\documentclass[]{emulateapj}

\begin{document}

\title{Gravitational settling of $^{22}$Ne  and white dwarf evolution}

\author{E. Garc\'{\i}a--Berro\altaffilmark{1}}
\affil{Departament de F\'\i sica Aplicada, 
       Escola Polit\`ecnica Superior de Castelldefels,
       Universitat Polit\`ecnica de Catalunya,  
       Av. del Canal Ol\'\i mpic, s/n,  
       08860 Castelldefels,  
       Spain}
\email{garcia@fa.upc.edu}
\author{L. G. Althaus\altaffilmark{2}}
\email{althaus@fcaglp.unlp.edu.ar} 
\author{A. H. C\'orsico\altaffilmark{2}}
\affil{Facultad de Ciencias Astron\'omicas y Geof\'{\i}sicas,
       Universidad  Nacional de  La Plata,  
       Paseo del  Bosque s/n,
       (1900) La Plata, 
       Argentina}
\email{acorsico@fcaglp.unlp.edu.ar}
\and
\author{J. Isern\altaffilmark{1}}
\affil{Institut de Ci\`encies de l'Espai, CSIC, 
       Facultat de Ci\`encies, 
       Campus UAB, 
       08193 Bellaterra, 
       Spain}
\email{isern@ieec.fcr.es}
\altaffiltext{1}{Institut d'Estudis Espacials de Catalunya, 
                 c/ Gran Capit\`{a} 2--4, 
                 08034 Barcelona, 
                 Spain}
\altaffiltext{2}{Member of the Carrera del Investigador Cient\'{\i}fico y 
                 Tecnol\'ogico, CONICET (IALP), Argentina}

\begin{abstract}
We  study  the effects  of  the  sedimentation  of the  trace  element
$^{22}$Ne in the  cooling of white dwarfs.  In  contrast with previous
studies ---  which adopted  a simplified treatment  of the  effects of
$^{22}$Ne  sedimentation ---  this is  done self-consistently  for the
first time, using an up-to-date stellar evolutionary code in which the
diffusion  equation is  coupled  with  the full  set  of equations  of
stellar evolution.   Due the large  neutron excess of  $^{22}$Ne, this
isotope rapidly sediments in the interior of the white dwarf. Although
we explore  a wide range  of parameters, we  find that using  the most
reasonable  assumptions concerning the  diffusion coefficient  and the
physical state of the white dwarf interior the delay introduced by the
ensuing  chemical  differentation  is   minor  for  a  typical  $0.6\,
M_{\sun}$  white dwarf.  For  more massive  white dwarfs,  say $M_{\rm
WD}\sim  1.0\,  M_{\sun}$, the  delay  turns  out  to be  considerably
larger.   These results are  in qualitatively  good accord  with those
obtained in  previous studies, but we  find that the  magnitude of the
delay introduced  by $^{22}$Ne  sedimentation was underestimated  by a
factor of $\sim 2$. We also  perform a preliminary study of the impact
of  $^{22}$Ne sedimentation  on the  white dwarf  luminosity function.
Finally, we  hypothesize as well  on the possibility of  detecting the
sedimentation  of  $^{22}$Ne  using  pulsating  white  dwarfs  in  the
appropriate  effective temperature  range  with accurately  determined
rates of change of the observed periods.
\end{abstract}


\section{Introduction}

White  dwarf stars  are  the  final evolutionary  stage  for the  vast
majority of  stars and, hence, they play  a key role in  our quest for
understanding  the  structure and  history  of  our Galaxy.   Standard
stellar  evolution theory  predicts  that most  white  dwarfs are  the
descendants  of  post-asymptotic  gianch   branch  stars  of  low  and
intermediate  masses  that  reach  the  hot  white  dwarf  stage  with
hydrogen-rich surface layers.  The basic inner structure expected in a
typical white dwarf consists of a degenerate core mostly composed of a
mixture  of  carbon and  oxygen  (the  ashes  of core  helium  burning
resulting from  the evolution of the progenitor)  plus some impurities
resulting from the original metal content of the progenitor star.  The
most important  of these impurities  is $^{22}$Ne, which  results from
helium  burning on  $^{14}$N  --- built  up  during the  CNO cycle  of
hydrogen burning ---  and reaches an abundance by  mass of $X_{\rm Ne}
\approx Z_{\rm CNO}\approx$ 0.02 in Population I stars.

An accurate determination of the  rate at which white dwarfs cool down
constitutes a  fundamental issue  and provides a  independent valuable
cosmic clock to determine ages of many Galactic populations, including
the disk (Winget  et al. 1987; Garc\'{\i}a-Berro et  al. 1988; Hernanz
et al. 1994)  and globular and open clusters (Richer  et al. 1997; Von
Hippel \& Gilmore 2000; Hansen et  al. 2002; Von Hippel et al.  2006).
Thus,  considerable   effort  has  been   devoted  to  observationally
determine the luminosity function of field white dwarfs --- which also
provides  a  measure  of  the  cooling rate  ---  and  to  empirically
determine  the  observed  white  dwarf  cooling  sequence  in  stellar
clusters. Additionally,  and from the  theoretical point of  view, the
development  of very  detailed white  dwarf evolutionary  models, that
incorporate  a complete description  of the  main energy  sources, has
also been a priority, since it allows a meaningful comparison with the
increasing wealth of observational data (Fontaine et al. 2001; Salaris
et al. 2000; Althaus \& Benvenuto 1998).

White dwarf evolution  can be basically described as  a simple cooling
process  (Mestel 1952)  in  which  the decrease  in  the thermal  heat
content of  the ions constitutes  the main source of  luminosity.  The
release of both latent heat (Van Horn 1968; Lamb \& Van Horn 1975) and
of gravitational energy due to a change in chemical composition during
crystallization  (Stevenson  1980;  Garc\'\i  a--Berro et  al.   1988;
Segretain  et  al.  1994;  Isern  et  al.  1991,  2000)  also  affects
considerably   the   cooling   of   white  dwarfs.    In   particular,
compositional separation at crystallization temporarily slows down the
cooling. This, in turn, influences  the position of the cut-off of the
disk white dwarf  luminosity function (Hernanz et al.  1994), which is
essential in obtaining and independent determination of the age of the
Galactic disk.

Another  potential source  of energy  are  the minor  species rich  in
neutrons,  such as  $^{22}$Ne.  Indeed,  as  first noted  by Bravo  et
al. (1992)  the two  extra neutrons present  in the  $^{22}$Ne nucleus
(relative to $A=2Z$) results in a net downward gravitational force and
a slow, diffusive settling of  $^{22}$Ne in the liquid regions towards
the center of the white  dwarf. The role of $^{22}$Ne sedimentation in
the  energetics of crystallizing  white dwarf  was first  addressed by
Isern et  al.  (1991).   The extent  to which the  cooling of  a white
dwarf  could be  modified by  this process  was later  investigated by
Bildsten  \& Hall  (2001)  and quantitatively  explored  by Deloye  \&
Bildsten  (2002).  Depending  on the  underlying uncertainties  in the
physics of the interdiffusion  coefficients, Deloye \& Bildsten (2002)
concluded that $^{22}$Ne sedimentation could release sufficient energy
to affect appreciably the cooling of massive white dwarfs, making them
appear bright  for very long periods  of time, of the  order of $10^9$
yr.  Deloye \&  Bildsten (2002) predicted that the  possible impact of
$^{22}$Ne  sedimentation  on white  dwarf  cooling  could  be seen  in
metal-rich clusters.  In fact, white dwarfs resulting from progenitors
formed in metal-rich systems are expected to have larger abundances of
$^{22}$Ne  in their cores,  and the  delay in  the cooling  history of
white  dwarfs  resulting from  $^{22}$Ne  diffusion  would be  largely
amplified in these clusters.

The recent  detection of a  markedly bright white dwarf  population in
the  old open  cluster  NGC 6791  (Bedin  et al.   2005) is  certainly
promising  in   this  regard.   Indeed,   the  significant  supersolar
metallicity characterizing  NGC 6791 (about [Fe/H]  $\sim +0.4$) turns
this cluster into a unique target to test the predictions of Deloye \&
Bildsten (2002), who claim that  the effects of sedimentation would be
largest in this cluster.   Although the anomalously bright white dwarf
luminosity  function observed  in  NGC 6791  could  be reflecting  the
presence of  a large  population of massive  white dwarfs  with helium
cores instead  of carbon-oxygen cores, as suggested  by Hansen (2005),
$^{22}$Ne   diffusion  constitutes  a   viable  explanation   for  the
morphology of the observed white dwarf luminosity function that cannot
be discarded (Kalirai et al.  2007).

In view  of these considerations,  a full and consistent  treatment of
white dwarf  evolution is  required to ascertain  the validity  of the
predictions of Deloye  \& Bildsten (2002) about the  role of $^{22}$Ne
diffusion in cooling white  dwarfs, since their findings were obtained
using a simplified treatment of white dwarf evolution.  In particular,
although  Deloye  \& Bildsten  (2002)  carried  out  a full  numerical
calculation  of  the  diffusion   process,  the  white  dwarf  cooling
sequences  were  computed  using   a  series  of  static  white  dwarf
mechanical configurations in which  the thermal evolution of the white
dwarf was decoupled and, moreover, the authors did not mention whether
or  not the  mechanical structure  of the  white dwarf  was recomputed
self-consistently at  each time step.   Another crucial point  is that
the white dwarf  models of Deloye \& Bildsten  (2002) were constructed
using  a  fully  degenerate  equation  of state  everywhere  (with  no
inclusion of  ion contributions  in the structure  calculation).  This
assumption is valid for the white dwarf interior, where the degeneracy
is high,  but not for the  outer nondegenerate parts of  the star.  In
addition, the  treatment of Deloye  \& Bildsten (2002) required  as an
additional input  a relationship between  the luminosity of  the white
dwarf and the  temperature of the core --- for  which they adopted the
results of  Althaus \& Benvenuto  (1998) --- which,  additionally, was
assumed to  be isothermal.   As a result,  the heat sources  and sinks
could  only be  evaluated  globally and  not  locally as  we do  here.
Consequently,  these  authors  could  not obtain  absolute  ages  but,
rather, relative delays  with respect to the absolute  ages derived by
Althaus  \&  Benvenuto  (1998).   The  aim of  the  present  work  is,
precisely, to fill  this gap.  Specifically, we present  a set of full
evolutionary and self-consistent  white dwarf cooling sequences, which
include  an accurate treatment  of time-dependent  $^{22}$Ne diffusion
and the physical state of  the white dwarf interior.  Special emphasis
is  given to the  treatment of  the luminosity  equation, particularly
regarding  the  contribution  to  the  energy  balance  of  the  terms
resulting  from  changes  in  the  profiles  of  chemical  abundances.
Calculations are followed down  to very low surface luminosities.  The
plan of the paper is  the following.  Section 2 contains details about
the input  physcis of our white  dwarf evolutionary code  and the main
aspects  of the  diffusion  treatment.   In Sect.   3  we present  our
results.  Finally,  Sect.  4 is devoted to  discussing and summarizing
our findings.

\section{Input of the models and evolutionary sequences}

\subsection{General description of the code}

In  this  work  we  have   computed  the  evolution  of  white  dwarfs
self-consistently, taking  into account  the changes in  the abundance
profiles induced by the  gravitational settling of $^{22}$Ne. That is,
we have solved the full  set of equations describing the structure and
evolution of  white dwarfs with the  luminosity equation appropriately
modified to account for the energy released by the changes in the core
abundances as a  consequence of $^{22}$Ne sedimentation. Consequently,
this  study constitutes  a  notable improvement  with  respect to  the
approach adopted  by Deloye  \& Bildsten (2002)  to compute  the white
dwarf  cooling sequences,  where all  energy sources  were  treated as
global quantities.

The evolutionary code employed in this study is the one we used in our
study of  the evolution of massive  white dwarf stars  (Althaus et al.
2007), modified to incorporate  the effects of $^{22}$Ne sedimentation
in the interior of white dwarfs.  In particular, the equation of state
for the  high-density regime is very  detailed and it  is described in
Segretain  et  al.   (1994).    It  accounts  for  all  the  important
contributions for  both the liquid  and solid phases.  The  release of
latent heat  upon crystallization ---  which is assumed to  occur when
the  ion  coupling constant  reaches  $\Gamma$=  180 (Stringfellow  et
al. 1990;  Chabrier 1993)  --- and neutrino  emission rates  have been
included following  Althaus et  al.  (2007).  Radiative  opacities are
those  from  OPAL  (Iglesias  \&  Rogers  1996)  complemented  at  low
temperatures   with  the  Alexander   \&  Ferguson   (1994)  molecular
opacities.   Chemical  redistribution  due  to phase  separation  upon
crystallization  has not  been  considered in  this work.   Particular
attention is  given to the treatment  of the very outer  layers of our
models at advanced stages of evolution. This is crucial in determining
absolute ages.  In particular, compression of these layers constitutes
the main  energy source of  the white dwarf  when it enters  the Debye
regime.  In  the present calculations,  the internal solutions  of the
Henyey iteration have been considered up to a fitting mass fraction of
$ \approx 10^{-14} M_r$.

\begin{figure}
\begin{center}
\includegraphics[clip,width=0.9\columnwidth]{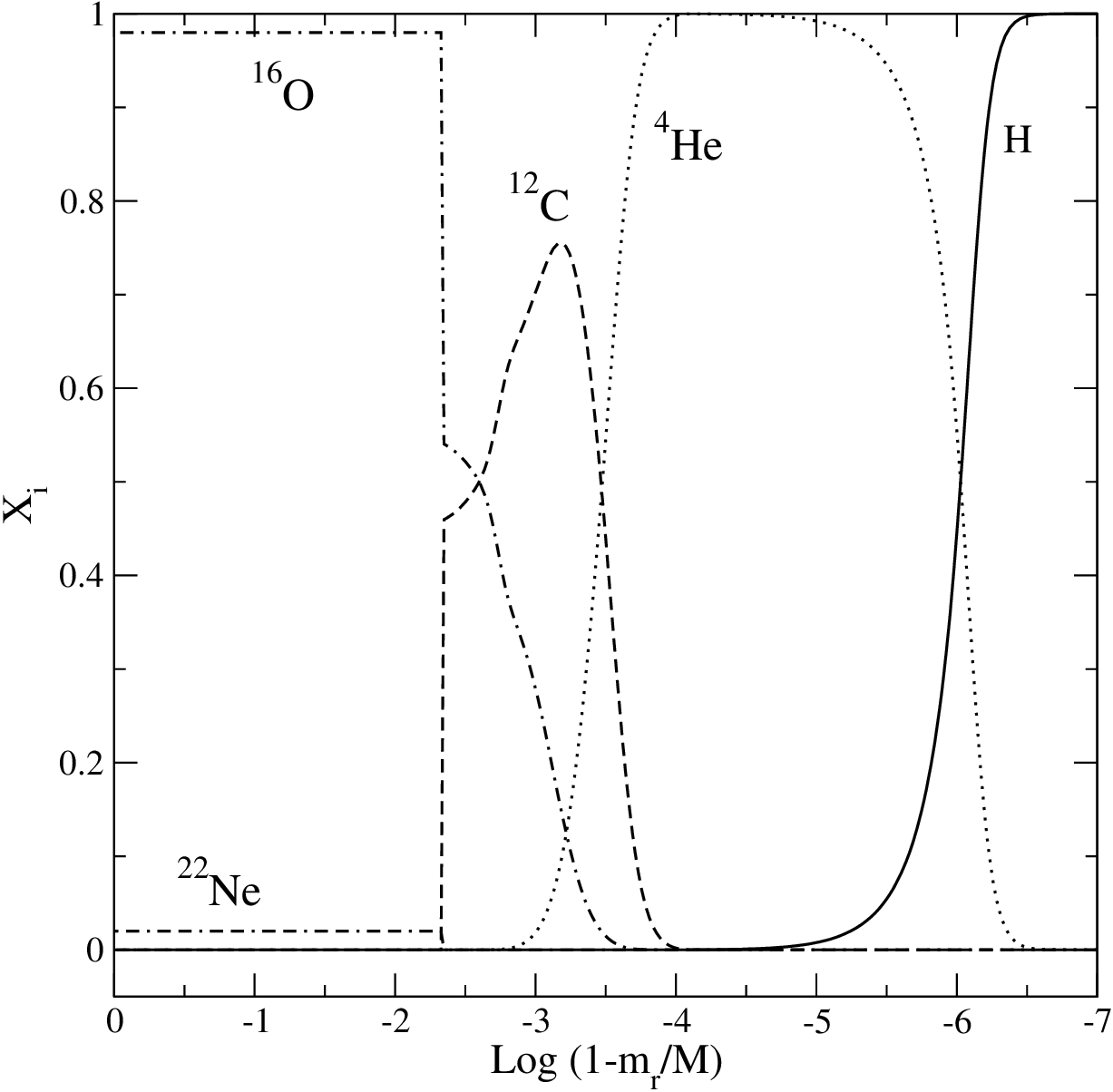}
\caption{The initial chemical abundance distribution for our $1.06  \,
         M_{\sun}$ oxygen-core white dwarf.}
\label{quimiini}
\end{center}
\end{figure}

The  initial  white dwarf  configuration  from  which  we started  our
calculations of  the cooling sequences  correspond to hot  white dwarf
structures  that were obtained  following the  artificial evolutionary
procedure described  in Althaus et  al (2005).  Because the  impact of
$^{22}$Ne diffusion  on the  cooling times is  irrelevant for  the hot
white  dwarf regime, the  precise election  of the  initial conditions
bears virtually no relevance for the purposes of the present work.  We
compute the evolution of white dwarf models with stellar masses of 0.6
and $1.06\, M_{\sun}$.  More massive white dwarfs are expected to have
cores composed mainly of oxygen and neon, the result of carbon burning
in semidegenerate  conditions in the  progenitor star (Ritossa  et al.
1996; Garc\'{\i}a--Berro et al.   1997).  The core chemical profile of
our models  consists of a predominant chemical  element (either carbon
or oxygen) plus $^{22}$Ne  with an initially flat abundance throughout
the core.  The  abundance of $^{22}$Ne results from  helium burning on
$^{14}$N     via     the     reactions    \mbox     {$^{14}$N($\alpha,
\gamma$)$^{18}$F($\beta^+$)$^{18}$O($\alpha,  \gamma$)$^{22}$Ne}.   In
this work,  we explore  the evolution of  white dwarfs  resulting from
Population  I star  progenitors,  i.e., characterized  by $X_{\rm  Ne}
\approx  Z_{\rm CNO}\approx$  0.02.   Although the  theory of  stellar
evolution  predicts the  core  of  most white  dwarfs  to be  composed
essentially  by a  mixture  of  carbon and  oxygen,  we assume  either
carbon-  or   oxygen-dominated  cores   to  be  consistent   with  the
one-component plasma  underlying assumption  in the derivation  of the
diffusion coefficient. The outer  layer chemical stratification of all
of our  models consists of a  pure hydrogen envelope  of $10^{-6} M_*$
(plus  a small  inner tail)  overlying a  helium-dominated  shell and,
below that, a buffer rich  in carbon and oxygen.  The initial chemical
stratification  corresponding  to an  oxygen-rich  $1.06 \,  M_{\sun}$
white dwarf  is displayed  in Fig. \ref{quimiini}.   The shape  of the
outer  layer chemical  profile is  given by  element diffusion  at low
luminosities.  Nevertheless,  diffusion occurring in  the outer layers
was switched off in the present calculations.  The evolutionary stages
computed cover the luminosity  range from $\log(L/L_{\sun}) \approx 0$
down to $-5.3$.

For a proper  treatment of white dwarf evolution  in a self-consistent
way  with the  changes in  the  core chemical  composition induced  by
$^{22}$Ne sedimentation, the standard luminosity equation for evolving
white dwarfs (thermonuclear reactions are neglected)

\begin{equation}
\frac{\partial L_r}{\partial M_r}= -\epsilon_\nu - C_P \dot T + 
{{\delta} \over {\rho}}\dot P,
\label{lumistandard}
\end{equation}

\noindent has to be appropriately modified --- see Isern et al. (1997)
and  also Kippenhahn  et al.   (1965).  In  Eq.  (\ref{lumistandard}),
$\epsilon_\nu$  denotes the  energy per  unit mass  per second  due to
neutrino  losses and  the other  quantities have  their  usual meaning
(Kippenhahn  \& Weigert  1990).   To  this end,  we  write the  energy
equation in terms of changes in the internal energy per gram ($u$) and
density ($\rho)$

\begin{equation}
\frac{\partial L_r}{\partial M_r}= -\epsilon_\nu - \dot u + {{P} \over 
{\rho^2}}\ \dot \rho,
\label{lumis1}
\end{equation}

\noindent and we assume  the white dwarf interior  to  be made of  two
chemical elements with abundance by  mass $X_1$ and $X_2$ ($X_1 + X_2$
= 1), where $X_1$ refers to the $^{22}$Ne abundance. Now,

\begin{equation}
\dot u= \left(\frac{\partial u}{\partial \rho} \right)_{T,X_1}\dot \rho\ +\
        \left(\frac{\partial u}{\partial T} \right)_{\rho,X_1}\dot T\ +\
	\left(\frac{\partial u}{\partial X_1} \right)_{\rho,T}\dot X_1 
\label{udot}
\end{equation}

\noindent and with the help of the thermodynamic relation

\begin{equation}
\left(\frac{\partial u}{\partial \rho} \right)_{T,X_1} = \ {{P} \over {\rho^2}}
        \ -\ {{T} \over {\rho^2}} \left(\frac{\partial P}{\partial T } 
        \right)_{\rho,X_1}, 
\label{urho}
\end{equation}

\noindent Eq.  (\ref{lumis1}) can be rewritten as

\begin{eqnarray}
\frac{\partial L_r}{\partial M_r}= && -\epsilon_\nu\ +\ {{T} \over {\rho^2}} 
\left(\frac{\partial P}{\partial T }\right)_{\rho,X_1} \dot \rho\ -\ C_V 
\dot T 
\nonumber\\
&& - \left(\frac{\partial u}{\partial X_1} \right)_{\rho,T} \dot X_1.
\label{lumisx}
\end{eqnarray}

Finally,  by  taking into account that $\rho=   \rho(P,T,X_1)$  and  the
definitions  

\begin{eqnarray}
\delta=&-&\left(\frac{\partial\ln\rho}{\partial\ln T}\right)_{P,X_1}\nonumber\\
\alpha=&\,&\left(\frac{\partial\ln\rho}{\partial\ln P}\right)_{T,X_1} \nonumber
\end{eqnarray}

\noindent Eq. (\ref{lumisx}) can be written as

\begin{equation}
\frac{\partial L_r}{\partial M_r}= - \epsilon_\nu - C_P \dot T + {{\delta} 
\over {\rho}} \dot P - A \dot X_1
\label{lumifin}
\end{equation}

\noindent where $A$ is given by

\begin{equation}
A= \left(\frac{\partial u}{\partial X_1} \right)_{\rho,T} -\ {{P \delta} 
\over {\alpha \rho^2}} \left(\frac{\partial \rho}{\partial X_1} \right)_{P,T},
\label{terminoA}
\end{equation}

\noindent or alternatively

\begin{equation}
A= \left(\frac{\partial u}{\partial X_1} \right)_{\rho,T} +\ {{\delta} \over 
{\rho}} \left(\frac{\partial P}{\partial X_1} \right)_{\rho,T}.
\label{terminoAfinal}
\end{equation}

The second and  third terms of Eq. (\ref{lumifin})  are the well-known
contributions of the  heat capacity and pressure changes  to the local
luminosity  of  the  star.   The  fourth term  represents  the  energy
released by chemical abundance  changes. Although this term is usually
small for  most stages of stellar  evolution --- as  compared with the
release of nuclear energy (Kippenhahn  et al.  1965) --- it will play,
as it will be clear below, a major role in the cooling of white dwarfs
with   diffusively   evolving   core   chemical   compositions.    For
neutron-rich species, such as in the problem of $^{22}$Ne diffusion in
the  core of  white dwarfs,  the  derivative $(\partial  u /  \partial
X_1)_{\rho, T}$  is dominated  by the electronic  contributions. Thus,
$A$ becomes negative and the  last term in Eq. (\ref{lumifin}) will be
a source (sink) of energy in those regions where diffusion leads to an
increase  (decrease) in  the  $^{22}$Ne local  abundance.  In  simpler
terms, an increase (decrease) in the $^{22}$Ne abundance --- or in the
molecular weight per electron $\mu_e$  --- will force the electron gas
to release (absorb) energy.   Finally, in Eq.  (\ref{lumifin}) we have
included a  term that  represents the contribution  of latent  heat of
crystallization ($\approx  k_{\rm B}T$ per ion, being  $k_{\rm B}$ the
Boltzmann constant).

\subsection{Treatment of diffusion}

The evolution  of the core  chemical abundance distribution  caused by
$^{22}$Ne  diffusion has  been fully  accounted for.   In  presence of
partial  gradients, gravitational and  electric forces,  the diffusion
velocities satisfy the set of equations (Burgers 1969)

\begin{eqnarray}
{{{\rm d}p_i} \over {{\rm d}r}}-{{\rho _i} \over \rho}{{{\rm d}p}
\over {{\rm d}r}}-n_iZ_ieE=
\sum\limits_{j\ne i}^{N} {K_{ij}}\left({w_j-w_i} \right). 
\label{burger}
\end{eqnarray}

\noindent  Here,  $p_i$,  $\rho_i$,   $n_i$,  $Z_i$  and  $w_i$  mean,
respectively,  the  partial pressure,  mass  density, number  density,
charge and diffusion velocity for species $i$ and $N$ means the number
of ionic species  plus electron.  The unknown variables  are $w_i$ and
the electric  field $E$.  The  resistence coefficients are  denoted by
$K_{ij}$.  For  the white dwarf core  we are mainly  interested in the
gravitational  settling,   thus  we  drop   off  the  first   term  in
Eq. (\ref{burger}). The  core of our white dwarf  models consists of a
predominant ion specie ---  either $^{16}$O or $^{12}$C plus $^{22}$Ne
(these two species  are denoted by subindices 2  and 1, respectively).
We assume the  electron to have zero mass.   Since, $\rho_i= A_i\ n_i\
m_p$, Eq. (\ref{burger}) can be written in the form

\begin{eqnarray}
A_1\ n_1\ m_p g - n_1\ Z_1 e E &=& K_{12}(\omega_2 - \omega_1) \\ A_2\
n_2\ m_p g - n_2\ Z_2 e E &=& K_{21}(\omega_1 - \omega_2) \\ A_1\ n_1\
\omega_1 + A_2\ n_2\ \omega_2 &=& 0
\end{eqnarray}

\noindent Here  $g$ is the local gravitational  acceleration, $m_p$ is
the proton mass,  and $A_i$ the atomic mass  number. The last equation
is the condition for no net  mass flow relative to the center of mass.
>From the last  set of equations we find  that the diffusion velocities
are (the effect of the induced electric field is considered)

\begin{eqnarray}
w_1&=& \left({{A_2}  \over{Z_2}} - {{A_1} \over{Z_1}}  \right) {{m_p g
D} \over{k_{\rm B} T}}\ { {\left({n_1 +  n_2} \right)} \over{n_1 n_2 f} } \\
w_2&=& - {{A_1 n_1} \over{A_2 n_2}} w_1,
\end{eqnarray}

\noindent where \mbox{$f=(1+A_1n_1/A_2n_2)(n_1Z_1+n_2Z_2)/n_1Z_1n_2Z_2$}, 
and we have introduced the diffusion coefficient $D$ as

\begin{eqnarray}
D= {{k_{\rm B} T n_1 n_2} \over{\left(n_1 + n_2 \right) K_{12} }}.
\end{eqnarray}

For $^{22}$Ne, $A_1= 22$ and $Z_1= 10$, while for either $^{16}$O  or  
$^{12}$C, $A_2/Z_2=2$. Thus, we arrive at

\begin{eqnarray}
w_1=& - &{{1} \over{5}} 
{{m_p g D} \over{k_{\rm B} T}}\ { {\left({n_1 + n_2} \right)} \over{n_1 n_2 f} } \\
w_2=&\, &  {{1} \over{5}} 
{{m_p g D} \over{k_{\rm B} T}}\ {A_1 \over{A_2}} { {\left({n_1 + n_2} \right)} 
\over{{n_2}^2 f} }. 
\end{eqnarray}

The negative  sign for the  $^{22}$Ne velocity reflects the  fact that
this  ion settles  towards the  center of  the white  dwarf.   This is
expected because of  the two excess neutrons of  the $^{22}$Ne nucleus
(in comparison with $A=2Z$). Note that if $^{22}$Ne is assumed to be a
trace element  ($n_1\approx 0$) then  \mbox{$(n_1+n_2)/n_1n_2f \approx
Z_1$} and we recover the diffusion velocity of Bildsten \& Hall (2001)
and Deloye \& Bildsten (2002), namely $w_{\rm Ne}= -2 m_p g D / k_{\rm
B} T$.

For the diffusion  coefficient $D$ we follow the  treatment adopted by
Deloye  \&  Bildsten  (2002)  for self-diffusion  coefficient  in  one
component plasmas.  In the  liquid interior $D$  becomes (in  units of
cm$^2$/s)
 
\begin{eqnarray}
D_s = { {7.3 \times 10^{-7} \ T } \over{\rho^{1/2} Z_2 \ 
\Gamma ^{1/3}}}, 
\label{difucoe}
\end{eqnarray}

\noindent where  $\Gamma$ is the  Coulomb coupling constant.   For the
regions  of  the white  dwarf  that  have  crystallized, diffusion  is
expected  to be  no longer  efficient due  to the  abrupt  increase in
viscosity expected  in the solid  phase.  Thus, we  set $D= 0$  in the
crystallized regions.   To avoid numerical difficulties,  the value of
$D$ is reduced from  $D_s$ to 0 in the solid phase  over a small range
of  $\Gamma$   values.   To   assess  the  consequences   of  possible
uncertainties  in the values  of $D$,  we have  performed evolutionary
calculations  also  for the  case  $D=  5D_s$.   Additionaly, we  have
explored the less likely situation  in which white dwarfs experience a
glassy transition at high $\Gamma$ values instead of crystallizing. In
a glassy state,  diffusion is expected to continue.   For this regime,
viscosity increases considerably with  the consequent reduction in the
diffusion coefficient.  Following also  Deloye \& Bildsten (2002), the
diffusion  coefficient  in the  glassy  state  is  reduced as  $D_{\rm
glass}=  D_s /  f^\star$  with \mbox  {$f^\star= 1+5.4\times  10^{-7}\
\Gamma^{2.3666} + 2.16 \times 10^{-27}\ \Gamma^{9.3666}$.}

The evolution  of the chemical  abundance distribution in the  core of
our white dwarf models caused by  diffusion is described in terms of a
time-dependent,   finite-difference  scheme   that   solves  elemental
continuity  equations for  the number  densities $n_i$.   The equation
governing the  evolution of the number  density of species  $i$ in the
space variable $r$ is

\begin{figure}
\begin{center}
\includegraphics[clip,width=0.9\columnwidth]{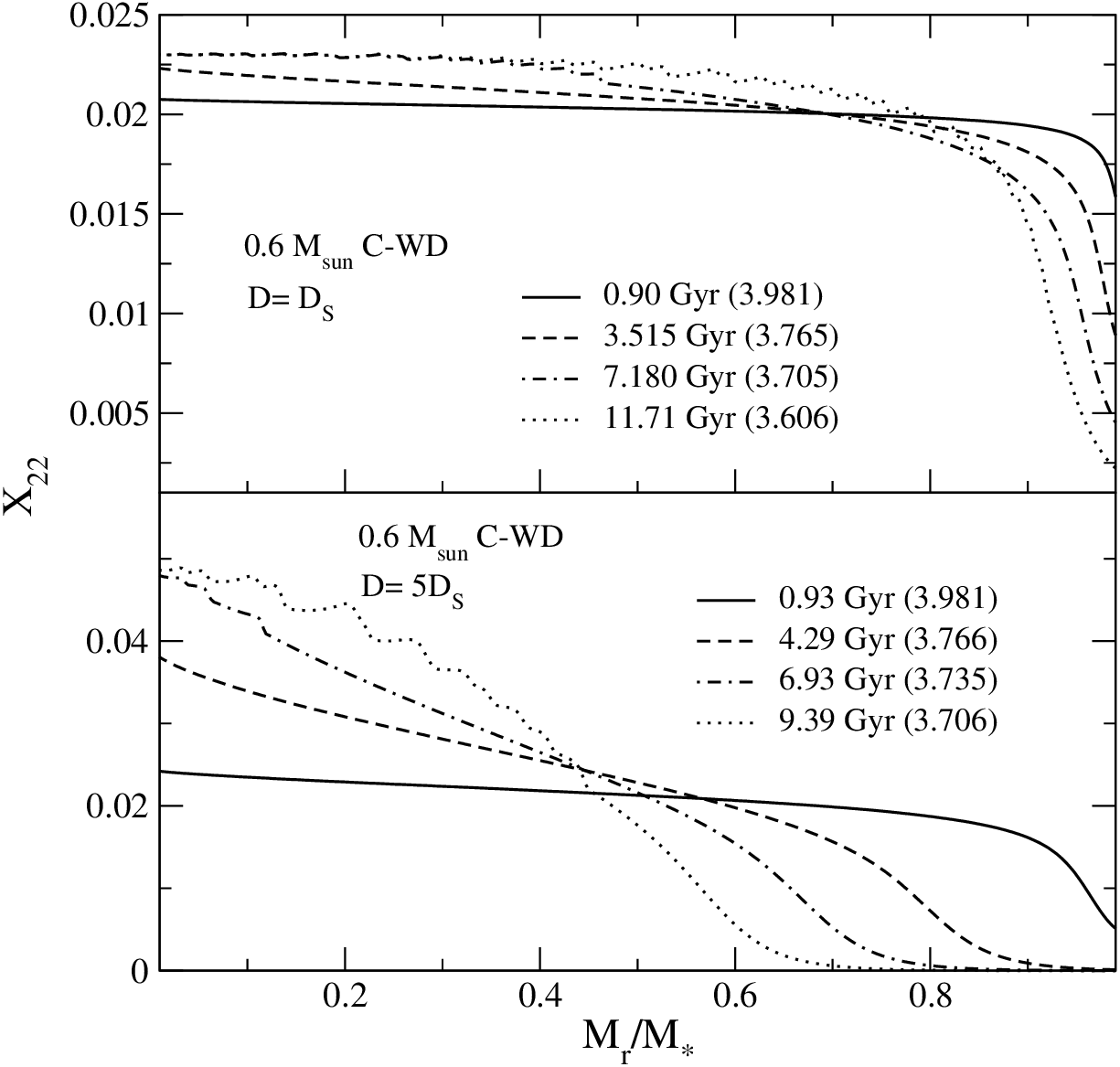}
\caption{Diffusively evolving internal $^{22}$Ne profiles for the $0.6
         \, M_{\sun}$ carbon-rich white dwarf sequence in terms of the
         mass fraction  at selected evolutionary stages.   The ages of
         the  models are  indicated  in the  legend.   The numbers  in
         brackets  are  the  values  of  $\log T_{\rm  eff}$  for  the
         corresponding models.  The  upper (bottom) panel displays the
         situation in  the case in which the  diffusion coefficient is
         set to $D= D_s$ ($D= 5 D_s$).}
\label{perfil_06}
\end{center}
\end{figure}

\begin{equation}
\frac{\partial n_i}{\partial t}= {{1} \over {r^2}}
\frac{\partial}{\partial r} \left( r^2 D 
\frac{\partial {n_i}}{\partial r} - r^2 \omega_i n_i \right), 
\label{fluxeq}
\end{equation}

\noindent supplemented with appropriate boundary conditions. Here, the
diffusive  flux  caused by   ion density  gradients and   the drifting
flux  due to  gravitational settling  are considered.   To  solve this
equation, we follow the  method described in Iben \& MacDonald (1985).  
In particular, we discretize Eq. (\ref{fluxeq}) as

\begin{eqnarray}
n_{i,k} - n_{i,k}^0= &&{{\Delta t} \over{\Delta V_k}} \Big[- 
\Big({r_{k+1/2}^2 \ \omega_{i,k+1/2}^0} \ {{n_{i,k} + n_{i,k+1} } \over{2}}
\nonumber\\  
&& -\ {r_{k-1/2}^2 \ \omega_{i,k-1/2}^0} \ {{n_{i,k} + n_{i,k-1} } \over{2}} 
\Big)\nonumber\\
&&  +\   {r_{k+1/2}^2  \  D_{k+1/2}^0}  \  {{n_{i,k+1}   -  n_{i,k}  }
\over{\Delta r_{k+1/2}}}\nonumber\\ 
&&  -\ {r_{k-1/2}^2 \ D_{k-1/2}^0}\ {{n_{i,k} - n_{i,k-1} } 
\over{\Delta r_{k-1/2}}}\Big],
\label{contin}
\end{eqnarray}

\noindent where the superscript $0$ means  that quantities  are to  be
evaluated  at  the beginning  of  the time  step  $\Delta  t$ and  the
subscript  $k$  refers  to  the  spatial  index.   \mbox{$\Delta  V_k=
(r_{k+1/2}^3- r_{k-1/2}^3)/3$} and $r_{k+1/2}^3= (r_k^3+r_{k+1}^3)/2$.
Note that $\omega_i$  and $D$ are also averaged  between adjacent grid
points.   Eq. (\ref{contin})  is  linear in  the  unknowns $n_i$.   We
follow the evolution of two chemical  species in the core of the white
dwarf:  either  $^{12}$C or  $^{16}$O,  and  $^{22}$Ne.   In order  to
compute the  dependence of  the structure and  evolution of  our white
dwarf models  on the varying abundances self-consistently,  the set of
equations describing  diffusion has  been coupled to  the evolutionary
code.

\section{Results}

\begin{figure}
\begin{center}
\includegraphics[clip,width=0.9\columnwidth]{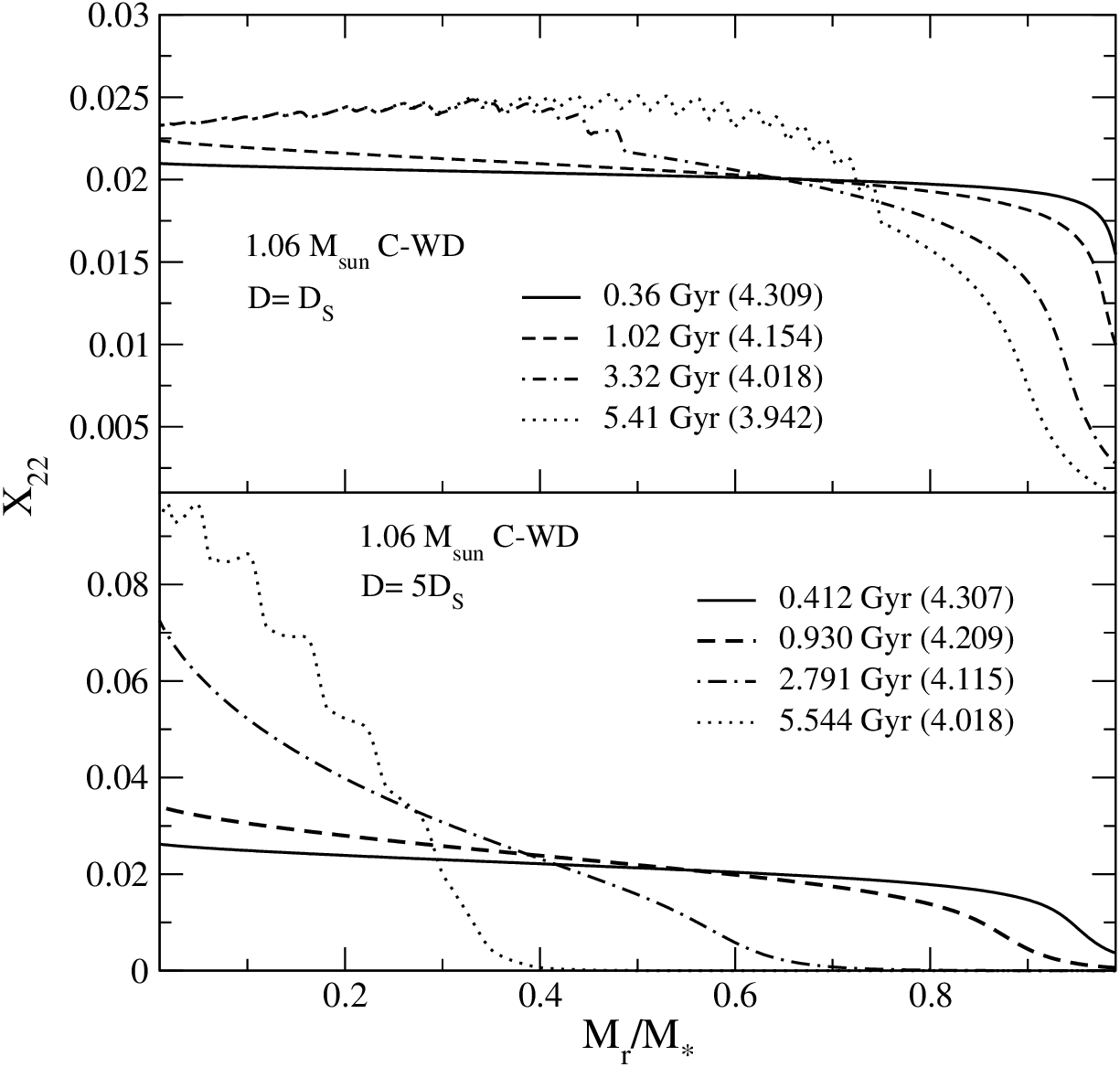}
\caption{Same as  Fig.  \ref{perfil_06} but for  the $1.06\, M_{\sun}$
         carbon-rich white dwarf models.}
\label{perfil_106}
\end{center}
\end{figure}

We  begin  by  examining  Fig.   \ref{perfil_06}  which  displays  the
diffusively  evolving $^{22}$Ne  profile  in the  core  of the  $0.6\,
M_{\sun}$ carbon-rich white  dwarf sequence as a function  of the mass
fraction.   The situation  at  four selected  stages  of evolution  is
depicted.   Clearly,  diffusion  appreciably  modifies  the  $^{22}$Ne
profile but only after long  enough time has elapsed, causing a strong
depletion of its abundance in the  outer region of the core, where the
diffusion timescale  is comparable to the  characteristic cooling time
of  the white  dwarf.   This is  more  apparent in  the  case of  very
efficient   diffusion,  as   illustrated  in   the  bottom   panel  of
Fig. \ref{perfil_06}, which shows the evolving profiles when $D= 5D_s$
is adopted.  In this case, by the time the white dwarf has started the
crystallization process  (at $\log T_{\rm eff}=  3.753$) diffusion has
completely depleted  $^{22}$Ne in the outermost region  (20\% by mass)
of  the  white dwarf,  and  has  markedly  enhanced the  abundance  of
$^{22}$Ne  in the  central regions  of the  star.   As crystallization
proceeds,  it  leaves recognizable  imprints  in  the inner  $^{22}$Ne
profile.   This   is  because  diffusion  is  not   operative  in  the
crystallized  region,  thus forcing  $^{22}$Ne  to  accumulate at  the
crystallization front.   As the crystallization  front moves outwards,
an apparently irregular profile is left behind, as clearly seen in the
figure.   The  apparent  wave   in  the  $^{22}$Ne  abundance  in  the
crystallized core is a numerical artifact --- see the discussion about
this issue in Deloye \& Bildsten (2002).  This irregular profile, that
arises  from   the  use  of  finite  timesteps   in  the  evolutionary
calculation and because diffusion  is switched-off at the liquid-solid
boundary which occurs over a finite range of $\Gamma$ values, bears no
relevance for the calculation of the cooling ages of the white dwarfs.

\begin{figure}
\begin{center}
\includegraphics[clip,width=0.9\columnwidth]{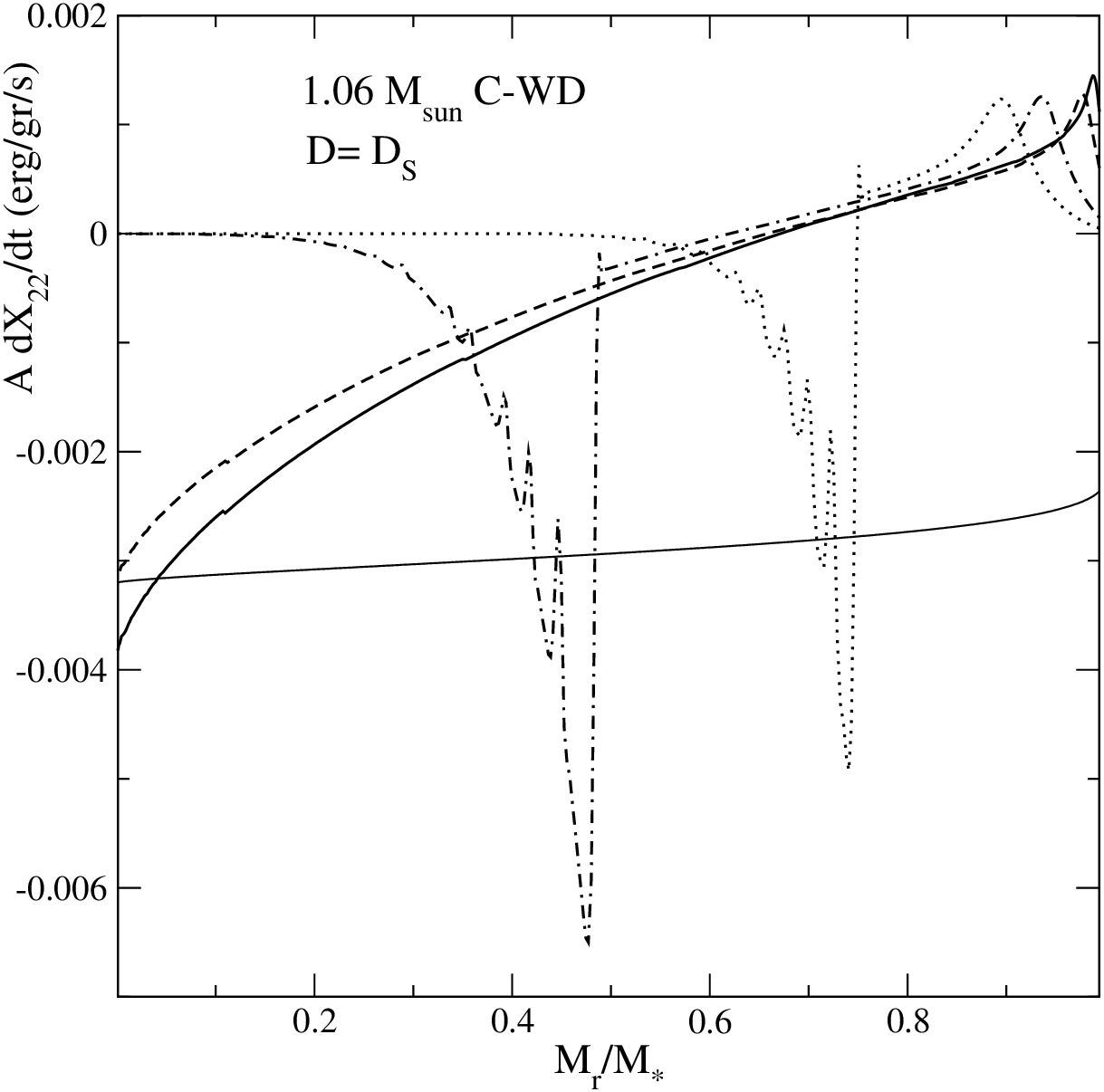}
\caption{The local  energy contribution resulting from  changes in the
         $^{22}$Ne  abundance  ---  the  term  $A  \dot  X_1$  in  Eq.
         (\ref{lumifin})   ---   for   the  same   $1.06\,   M_{\sun}$
         carbon-rich white  dwarf models shown  in the upper  panel of
         Fig.   \ref{perfil_106}.   The  thin  line denotes  the  heat
         capacity contribution  for the model with  $\log T_{\rm eff}=
         4.154$.   Note  that  in  the  outer  regions  of  the  core,
         $^{22}$Ne diffusion constitutes an energy sink ($A \dot X_1 >
         0$).}
\label{energia_106}
\end{center}
\end{figure}

The rate at which $^{22}$Ne diffuses downwards increases markedly with
gravity.   Thus, we  expect a  more rapid  sedimentation and  a faster
depletion of  $^{22}$Ne in the  outer layers of massive  white dwarfs.
This  is borne  out by  Fig. \ref{perfil_106},  which  illustrates the
$^{22}$Ne profile evolution for various $1.06 \, M_{\sun}$ carbon-rich
white dwarf  models. The  results for $D=  5D_s$, shown in  the bottom
panel of this figure, are  worthy of comment.  Indeed, while the white
dwarf  remains in  a  liquid state  (at  effective temperatures  $\log
T_{\rm  eff} > 4.12$),  the $^{22}$Ne  abundance distribution  will be
strongly modified by diffusion during the evolution.  Note that by the
time the white dwarf begins  to crystallize, $^{22}$Ne has diffused so
deep into  the core that no trace  of this element is  found in layers
even as  deep as  $0.4 \, M_{\sun}$  below the stellar  surface.  Note
also the marked accumulation of $^{22}$Ne towards the central regions.
Shortly  after, at  $\log T_{\rm  eff} \approx  4.02$, about  half the
white dwarf  mass is crystallized and the  $^{22}$Ne distribution will
remain  frozen with  further  cooling.  Because  massive white  dwarfs
crystallize earlier  than less massive ones,  $^{22}$Ne diffusion will
end  at much  higher effective  temperatures when  compared  with less
massive white dwarfs.  This is  a critical issue regarding the cooling
ages of massive white dwarfs, as it will be discussed below.

It is  clear from these  figures that diffusion  substantially changes
the core  abundance distribution of  $^{22}$Ne in the course  of white
dwarf  evolution.  These  changes  are in  good  agreement with  those
reported by Deloye  \& Bildsten (2002). We thus  would expect that the
contribution to  the local energy  budget of the white  dwarf stemming
from the  last term  in Eq. (\ref{lumifin})  could play  a significant
role.  We show  this in Fig.  \ref{energia_106} for  the same $1.06 \,
M_{\sun}$  carbon-rich stellar  models  shown in  the  upper panel  of
Fig. \ref{perfil_106}.   The figure clearly illustrates  the fact that
$^{22}$Ne  diffusion constitutes  a  local source  or  sink of  energy
depending  on whether the  local abundance  of $^{22}$Ne  increases or
decreases. In  fact, note that  for those regions where  diffusion has
increased  the local  abundance of  $^{22}$Ne,  $A \dot  X_{22} <  0$.
Thus, this  term constitutes an energy  source in such  regions --- in
fact, the increase in the molecular weight $\mu_e$ forces the electron
gas to  release energy.   In the outermost  part of the  core however,
where $^{22}$Ne  depletion occurs, $A \dot  X_1 > 0$,  thus implying a
sink of  energy in  those regions. To  compare with the  main standard
contribution to  $L_r$, we include in Fig.   \ref{energia_106} the run
of $C_p \dot T$ --- second  term in Eq.  (\ref{lumifin}) --- as a thin
solid line.  We do this for  the model with $\log T_{\rm eff}= 4.154$,
which corresponds to the model shown with dashed line.  We have chosen
this model because it corresponds to an evolutionary stage just before
the onset of crystallization.  For this model, the energy contribution
from $^{22}$Ne diffusion near the center of the white dwarf is similar
to  the  heat  capacity  contribution  ---  of  about  0.003  erg/g/s.
Finally, the dot-dashed and dotted curves correspond, respectively, to
models for which  50 and 75\% of their  mass has already crystallized.
The  $^{22}$Ne accumulation at  the crystallization  front leads  to a
large  local energy  contribution, restricted  to a  very  narrow mass
range.   Because diffusion  stops  in the  crystallized interior,  the
$^{22}$Ne  profile remains  frozen there,  and the  term $A  \dot X_1$
vanishes.  This  fact limits the  extent to which  $^{22}$Ne diffusion
constitutes an energy source for the star.

\begin{figure}
\begin{center}
\includegraphics[clip,width=0.9\columnwidth]{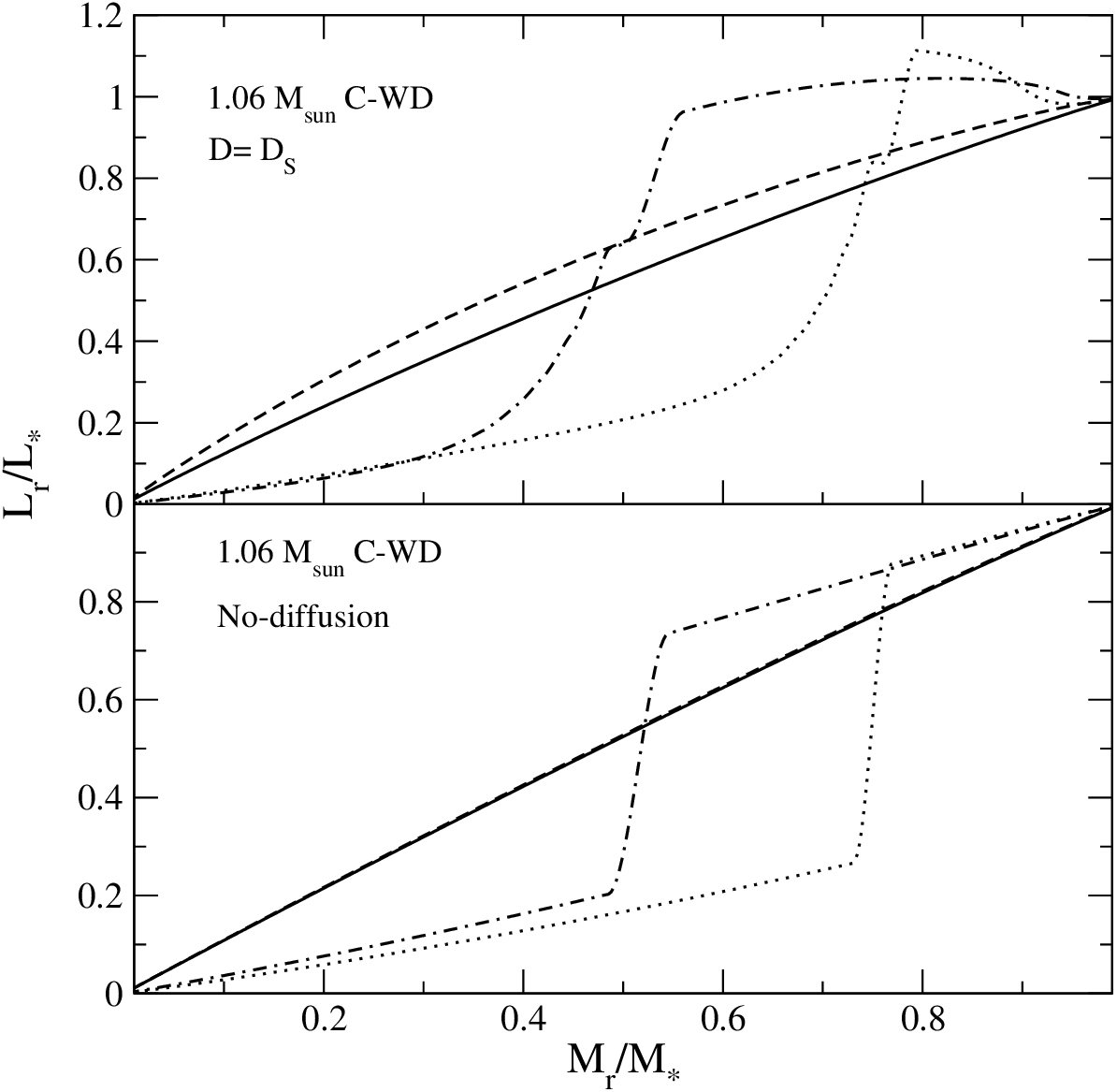}
\caption{Upper  panel:  The  fractional  luminosity in  terms  of  the
         fractional mass  for the same $1.06  \, M_{\sun}$ carbon-rich
         white  dwarf  models  shown   in  the  upper  panel  of  Fig.
         \ref{perfil_106}.  The  bottom panel shows  the same quantity
         in the case in which $^{22}$Ne diffusion is not considered.}
\label{lumifrac}
\end{center}
\end{figure}

\begin{figure}
\begin{center}
\includegraphics[clip,width=0.9\columnwidth]{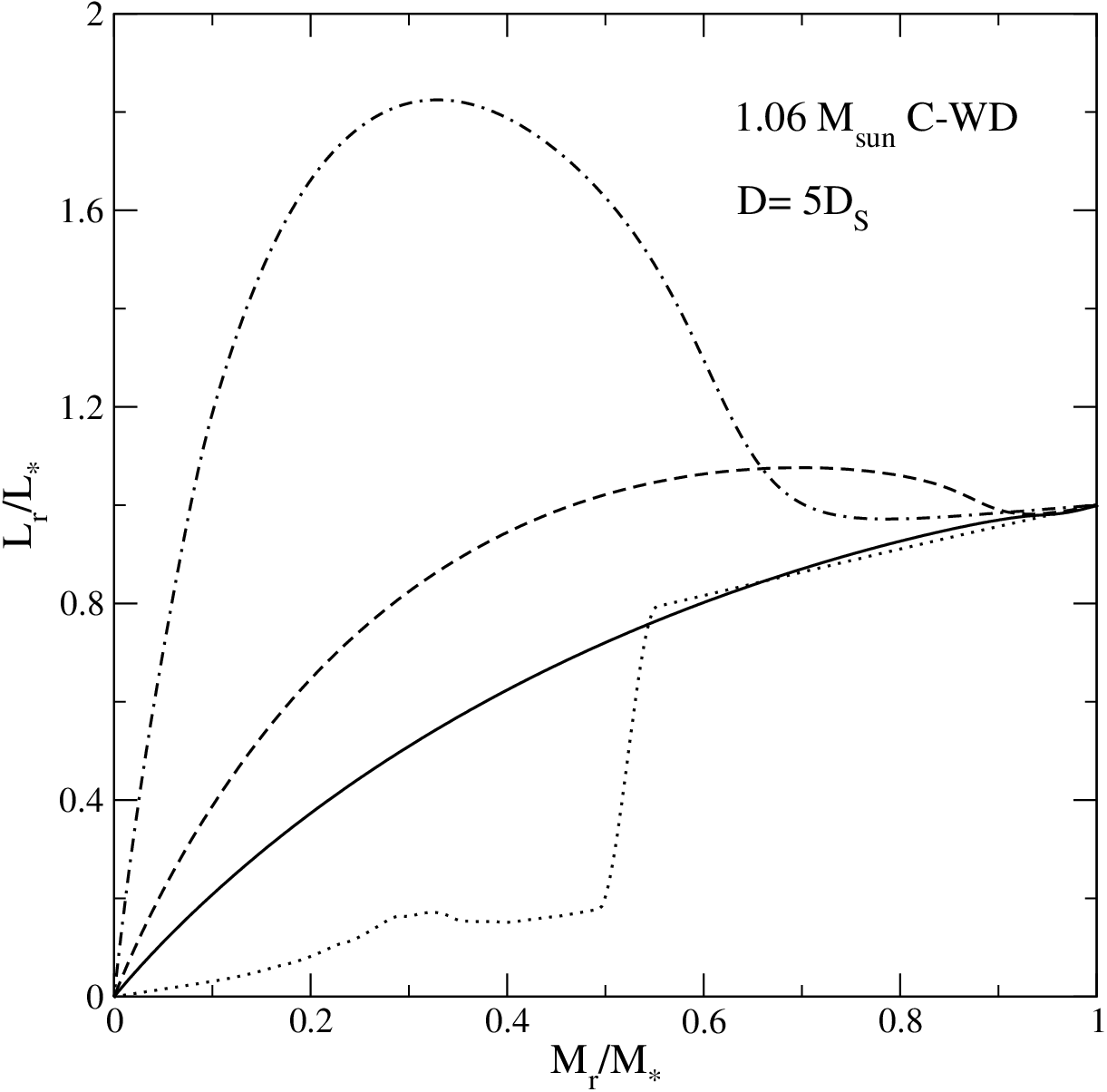}
\caption{The fractional luminosity in terms of the fractional mass for
         the  same $1.06\,  M_{\sun}$ carbon-rich  white  dwarf models
         shown in  the bottom panel of Fig.   \ref{perfil_106}, in the
         case in which we adopt $D=5 D_s$.}
\label{lumifrac_5ds}
\end{center}
\end{figure}

>From the above  discussion, it is expected that  the $^{22}$Ne profile
evolution affects the fractional  luminosity of the white dwarfs. This
is exemplified in the upper  panel of Fig. \ref{lumifrac}, which shows
the fractional luminosity $L_r$ in terms of the fractional mass, $M_r$
for the same $1.06\,  M_{\sun}$ carbon-rich white dwarf stellar models
shown  in  Fig.  \ref{energia_106}.   The  bottom  panel displays  the
behavior for the case in  which $^{22}$Ne diffusion is not considered.
In  both  cases,   the  imprint  of  the  latent   heat  release  upon
crystallization  is apparent  in the  models with  $\log  T_{\rm eff}=
4.018$ and  3.942. There is  clearly an important contribution  of the
$^{22}$Ne diffusion to the  luminosity budget of massive white dwarfs.
This  contribution  is  notably   enhanced  when  a  larger  diffusion
coefficient   is   considered,  as   it   becomes   clear  from   Fig.
\ref{lumifrac_5ds}, which shows the same quantity in the case in which
$D= 5D_s$ is adopted.  The models  plotted in this figure are the same
already  shown in the  bottom panel  of Fig.   \ref{perfil_106}.  Note
that the fractional luminosity  profile is indeed strongly modified in
this case.   It is particularly  relevant the model with  $\log T_{\rm
eff}= 4.115$ (dot-dashed line),  which has already a small crystalline
core.  In fact,  its internal luminosity at $M_r  \approx 0.3 M_\star$
is about  twice the luminosity  emerging from the surface.   Note also
the strong decrease  in the fractional luminosity in  the outer layers
of this model, which results  from the depletion of $^{22}$Ne in these
layers --- see the bottom panel of Fig.  \ref{perfil_106} --- with the
consequent energy absorption.

\begin{figure}
\begin{center}
\includegraphics[clip,width=0.9\columnwidth]{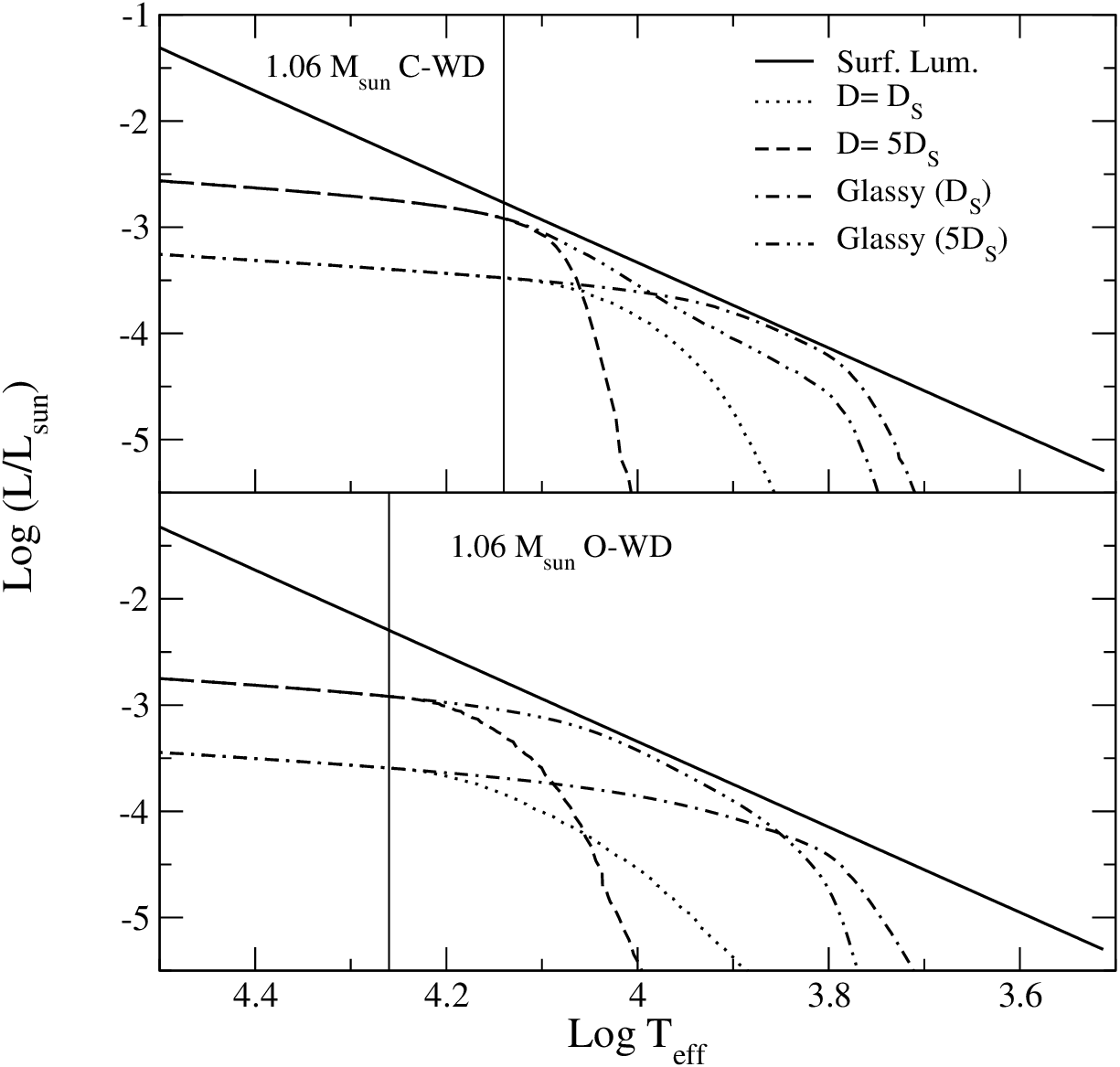}
\caption{Luminosity  contribution   in  solar  units   resulting  from
         $^{22}$Ne  diffusion in  terms of  the white  dwarf effective
         temperature for $1.06 \, M_{\sun}$ sequences with carbon- and
         oxygen-rich  cores (upper  and  bottom panel,  respectively).
         Dotted and dashed lines are  the results for $D= D_s$ and $D=
         5D_s$,  while the  remaining curves  correspond to  the cases
         when models enter a  glassy state.  For the latter situation,
         we  show  results for  $D=  D_{\rm  glass}$  and $D=  5D_{\rm
         glass}$.   The solid  line displays  the  surface luminosity.
         The  vertical  line  marks   the  approximate  value  of  the
         effective temperature for the onset of core crystallization.}
\label{hr106}
\end{center}
\end{figure}

To get a deeper insight  of the importance of $^{22}$Ne diffusion into
the global energetics during the  whole white dwarf evolution, we show
in Fig.  \ref{hr106} the resulting luminosity contribution obtained by
integrating at  each effective temperature  the term $-A \dot  X_1$ in
Eq.  (\ref{lumifin})  throughout all the star.   The result (expressed
in  solar  units) is  shown  in terms  of  the  white dwarf  effective
temperature for  the $1.06 \, M_{\sun}$ cooling  sequences.  The upper
panel corresponds to white dwarf sequences with carbon-rich cores.  To
highlight the role of the  core chemical composition, the bottom panel
shows the  sequences with oxygen-rich cores.   Several aspects deserve
comments.   To  begin  with,   note  that,  because  the  high  photon
luminosity  of the  star, the  luminosity contribution  from $^{22}$Ne
sedimentation is of a very minor importance during the hot white dwarf
stages.  It is only after  the onset of core crystallization that this
energy source contributes appreciably  to the star luminosity. This is
particularly  true   for  the   case  of  carbon-rich   models,  which
cyrstallize  at  lower  surface  luminosities than  their  oxygen-rich
counterparts.   Because   the  rapid  crystallization   of  the  core,
$^{22}$Ne   diffusion  luminosity   declines   steeply  with   further
cooling. This decline occurs earlier  in the case $D= 5D_s$ because of
the rapid depletion of $^{22}$Ne in the outer layers in this case (see
Fig. \ref{perfil_106}).  A similar  behavior was reported by Deloye \&
Bildsten   (2002).    As   illustrated   in  the   bottom   panel   of
Fig. \ref{hr106},  the impact of $^{22}$Ne settling  in the energetics
of white  dwarfs with oxygen-rich cores  is less relevant  than in the
case of  carbon--rich ones.  Finally,  for both core  compositions, we
compute the  $^{22}$Ne settling luminosity  for the case in  which the
white dwarf is assumed to experience a transition to a glassy state at
high $\Gamma$ values  instead of crystallizing.  We show  this for the
cases  $D= D_{\rm  glass}$  and $D=  5D_{\rm  glass}$ (dot-dashed  and
dot-dot-dashed lines,  respectively).  Because diffusion  is operative
in a glassy state, $^{22}$Ne settling luminosity obviously persists in
more  advanced stages  of the  evolution, providing  the bulk  of star
luminosity at small effective  temperatures. This is particularly true
for  the case  $D= 5D_{\rm  glass}$.  However,  at very  low effective
temperatures,  the viscosity  increase  in the  glassy state  prevents
diffusion from  occurring, and the  luminosity contribution eventually
declines.

\begin{figure}
\begin{center}
\includegraphics[clip,width=0.9\columnwidth]{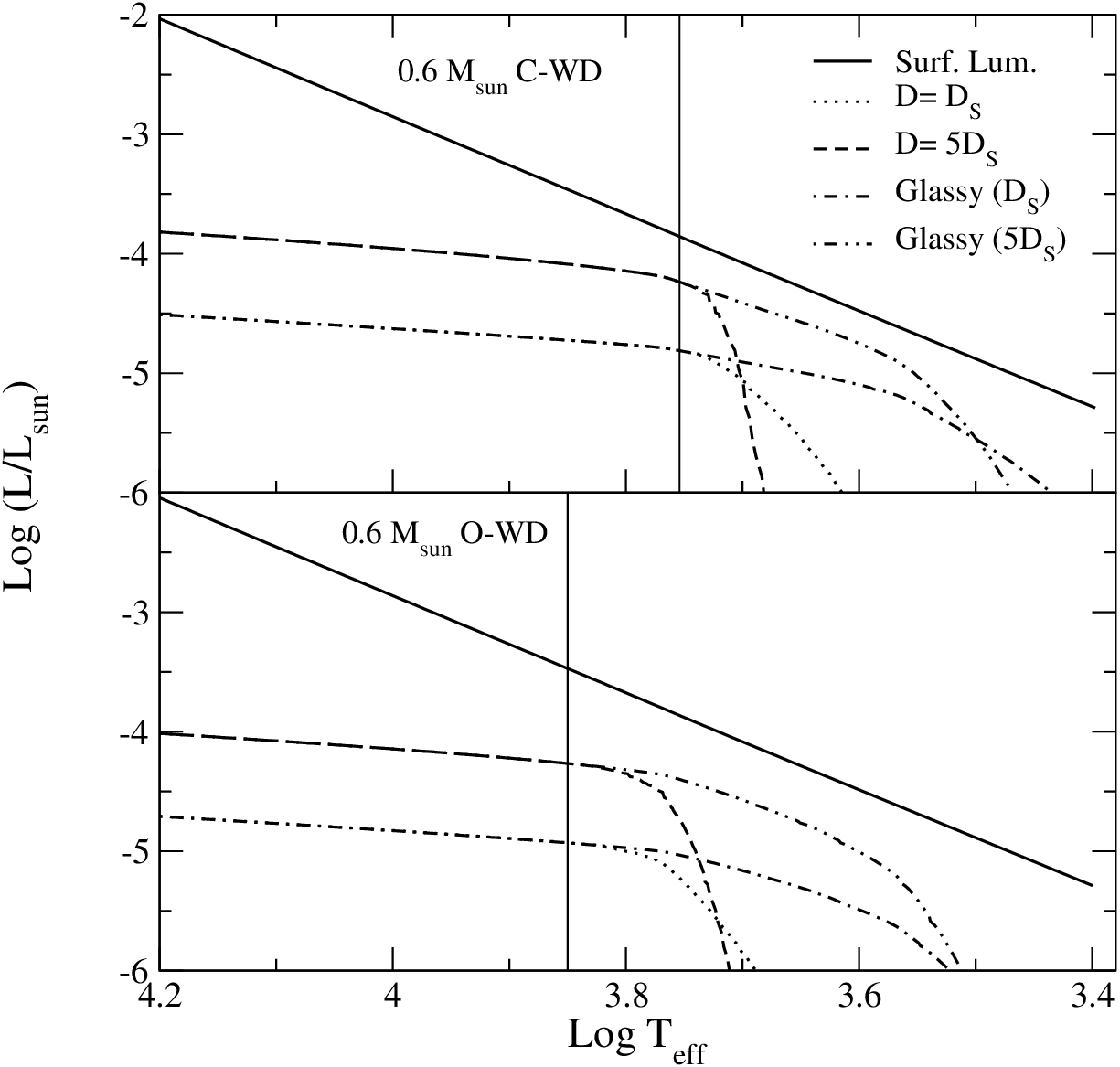}
\caption{Same as Fig. \ref{hr106} but  for the $0.6 \, M_{\sun}$ white
         dwarf sequences.}
\label{hr06}
\end{center}
\end{figure}

\begin{figure}
\begin{center}
\includegraphics[clip,width=0.9\columnwidth]{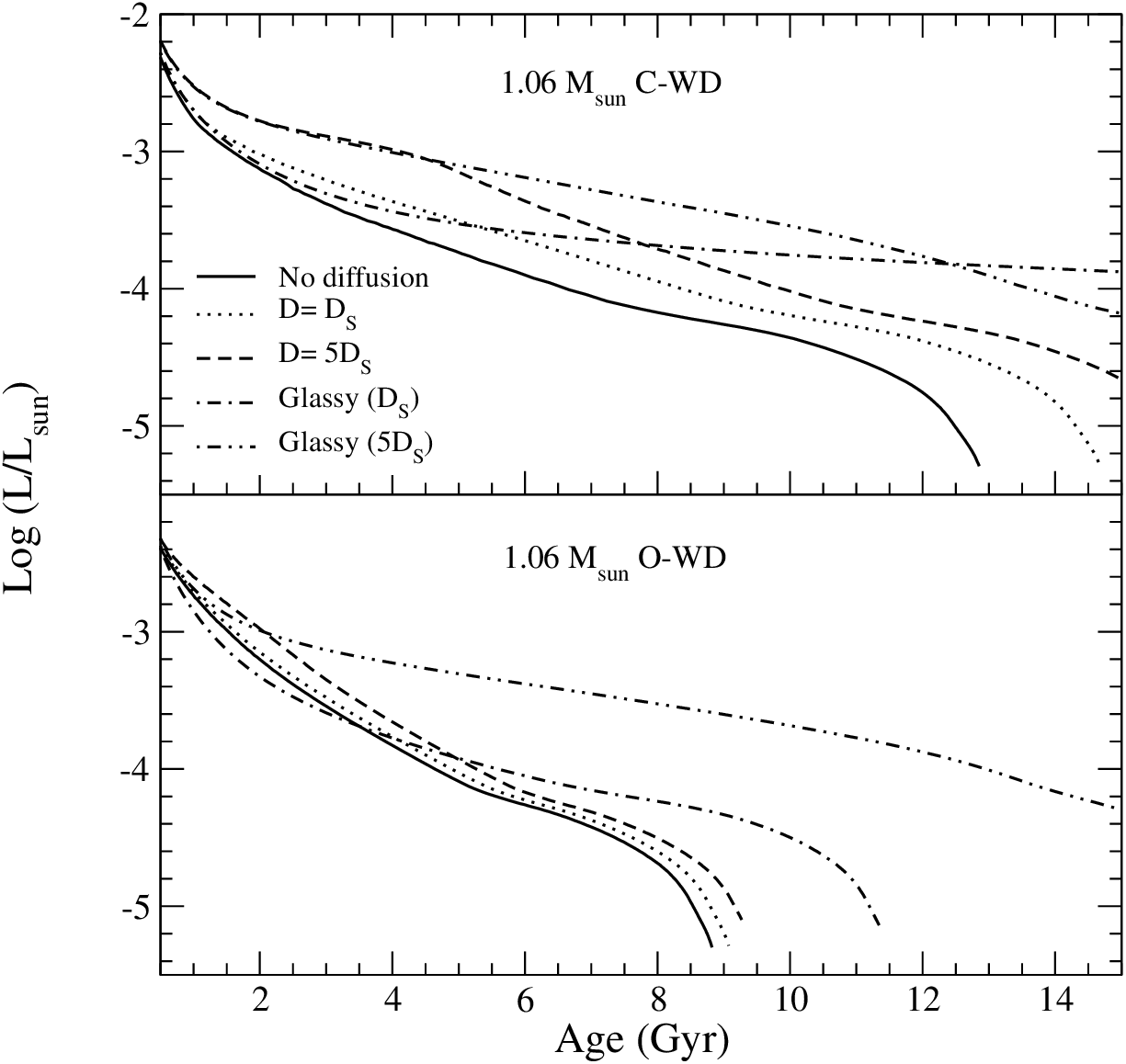}
\caption{Surface  luminosity versus  age  for the  $1.06 \,  M_{\sun}$
         white  dwarf  sequences with  carbon-  and oxygen-rich  cores
         (upper  and  bottom  panel,  respectively).  The  solid  line
         displays the  cooling times for  the case in  which $^{22}$Ne
         diffusion is not considered.  Dotted and dashed lines display
         the results for  $D= D_s$ and $D= 5D_s$,  while the remaining
         curves  correspond to the  cases when  models enter  a glassy
         state. For the latter situation, we show results for 
         $D= D_{\rm glass}$  and $D= 5D_{\rm glass}$.}
\label{edad106}
\end{center}
\end{figure}

For  low-mass  white  dwarfs,  the  impact of  $^{22}$Ne  settling  is
expectedly much  less noticeable, albeit  not negligible, particularly
in the case  of carbon-rich white dwarfs and  efficient diffusion (see
Fig.   \ref{hr06}).  Note  that because  $^{22}$Ne sedimentation  is a
slower  process  in low-mass  white  dwarfs,  it  will take  long  for
$^{22}$Ne settling  to contribute  to the energy  budget of  the star.
Since less  massive white dwarfs  crystallize at much  lower effective
temperatures than massive ones,  this contribution will play some role
at the  very late stages  of evolution.  It  is worth noting  that the
qualitative behavior of the evolution  of the global energetics of our
full sequences  resembles that reported by Deloye  \& Bildsten (2002).
In part,  this is expected  since --- as  noted by Deloye  \& Bildsten
(2002) --- the  diffusion timescale in  the white  dwarf core  is much
longer than the time on  which energy is transferred through the core.
We expect then  a similar behavior between both  sets of calculations,
as discussed below.
 
\begin{figure}
\begin{center}
\includegraphics[clip,width=0.9\columnwidth]{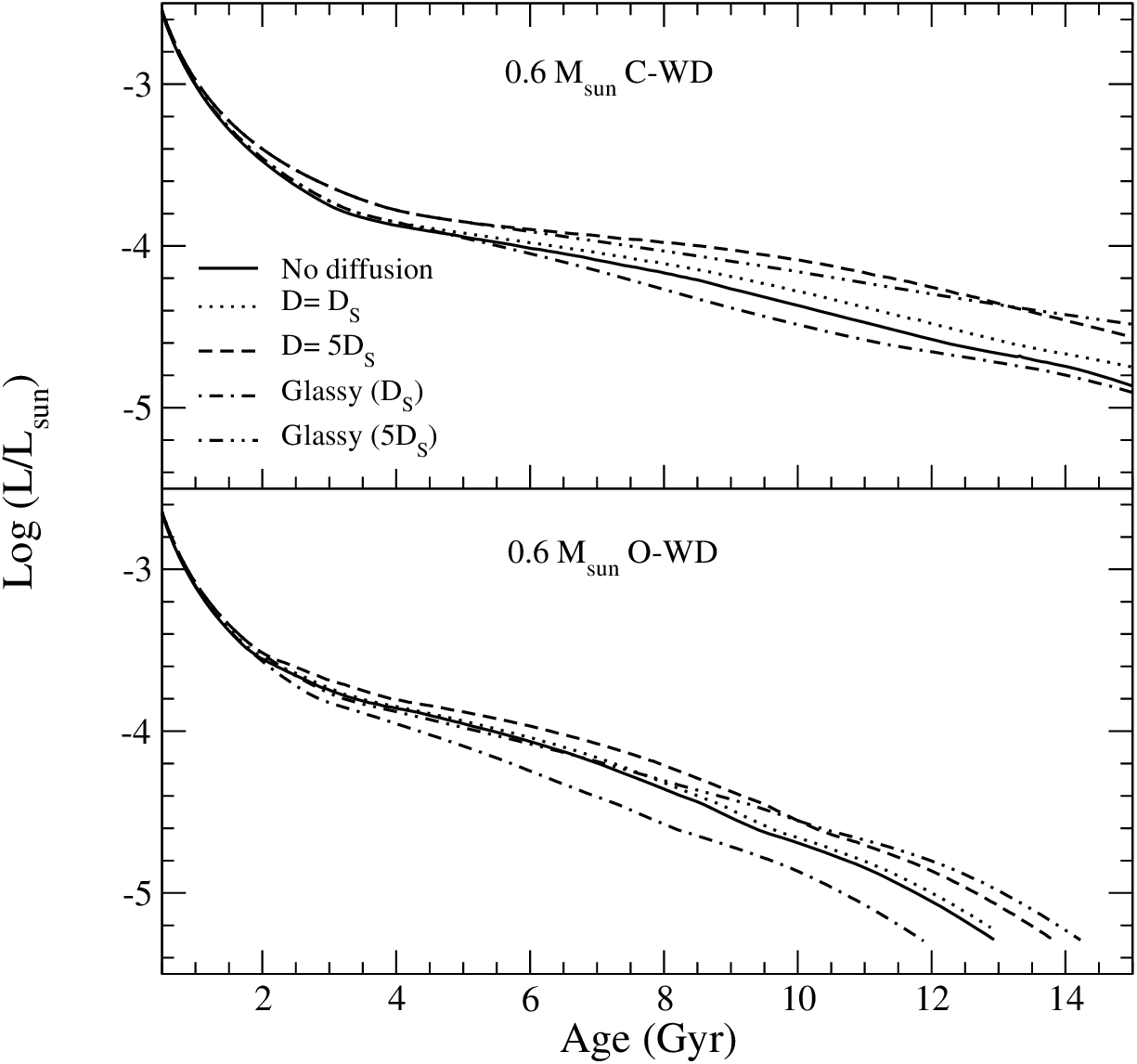}
\caption{Same  as Fig.  \ref{edad106} but  for the  $0.6  \, M_{\sun}$
         white dwarf sequences.}
\label{edad06}
\end{center}
\end{figure}

The impact of $^{22}$Ne sedimentation  in the white dwarf cooling ages
is seen in Figs.  \ref{edad106} and \ref{edad06} for the 1.06 and $0.6
\, M_{\sun}$  white dwarf cooling sequences,  respectively.  Here, the
white dwarf surface luminosity is shown as a function of the age.  The
solid line  corresponds to the standard case  when $^{22}$Ne diffusion
is not considered.  The top and bottom panels of each of these figures
are  for  carbon-   and  oxygen-rich  cores,  respectively.   Clearly,
$^{22}$Ne   diffusion  profoundly   influences   the  cooling   times,
particularly those  of massive  white dwarfs.  This  influence becomes
more dramatic in the case of efficient diffusion --- this is, the case
in  which $D= 5D_s$  is adopted  --- which  corresponds to  the dashed
curves.  The  star will  spend a long  time to  get rid of  the energy
released  by   the  diffusion-induced  abundance   changes,  with  the
consequent marked  lengthening of evolutionary  times persisting until
low luminosities.  From the  discussion in the preceding paragraph, it
is evident  that the  signatures of $^{22}$Ne  diffusion in  the white
dwarf  cooling track  start  to manifest  themselves  earlier in  more
massive  white dwarfs  ---  at $\log({L/L_{\sun}})  \approx -2$.   The
delay in the  cooling times depends not only on  the stellar mass, but
also markedly on the chemical  composition of the core.  For instance,
for  the  case  $D= 5D_s$,  the  delays  introduced  in the  $1.06  \,
M_{\sun}$  carbon-rich  sequence  amount  to  2.45  and  3.24  Gyr  at
$\log({L/L_{\sun}})  \approx -3$ and  $-4$, respectively,  as compared
with the 0.5 and 0.6 Gyr  in the case of the oxygen-rich sequence.  In
low-mass white  dwarfs, appreciable delays  in the cooling  rates take
place but only at low luminosities and for efficient diffusion.

The cooling rate in the case that white dwarfs experience a transition
to a  glassy state  behaves differently since  in this  case $^{22}$Ne
sedimentation continues  releasing energy even at  very large $\Gamma$
values. This distinct  behavior can be appreciated in  the two cooling
curves of the $1.06 \, M_{\sun}$ carbon-rich sequences shown as dashed
and  dot--dot--dashed  curves  which  correspond, respectively,  to  a
sequence that  experiences crystallization (with  $D= 5D_s$) and  to a
sequence  that  undergoes  a   glassy  transition  (with  $D=  5D_{\rm
glass}$).  In  fact, the decline  in the luminosity  contribution from
$^{22}$Ne sedimentation  due to the presence of  the crystalline core,
see Fig. \ref{hr106}, translates into a change of slope in the cooling
curve with $D= 5D_s$ at $\log({L/L_{\sun}}) \approx -3$.  By contrast,
the cooling  rate for  glassy white dwarfs  remains much  smaller.  We
thus  expect substantially  larger delays  in  the cooling  ages if  a
transition to a glassy state  actually occurs in nature. In this case,
massive  white dwarfs would  remain bright  even at  exceedingly large
ages.   Note that  during  the crystallization  stage, sequences  that
experience a  glassy transition (in  the case $D= D_{\rm  glass}$) are
younger than sequences without  diffusion.  This is because the glassy
white dwarf  does not release latent heat.  Because crystallization is
more  relevant in  less massive  white dwarfs,  this behavior  is more
evident in such white dwarfs.

\begin{figure}
\begin{center}
\includegraphics[clip,width=0.9\columnwidth]{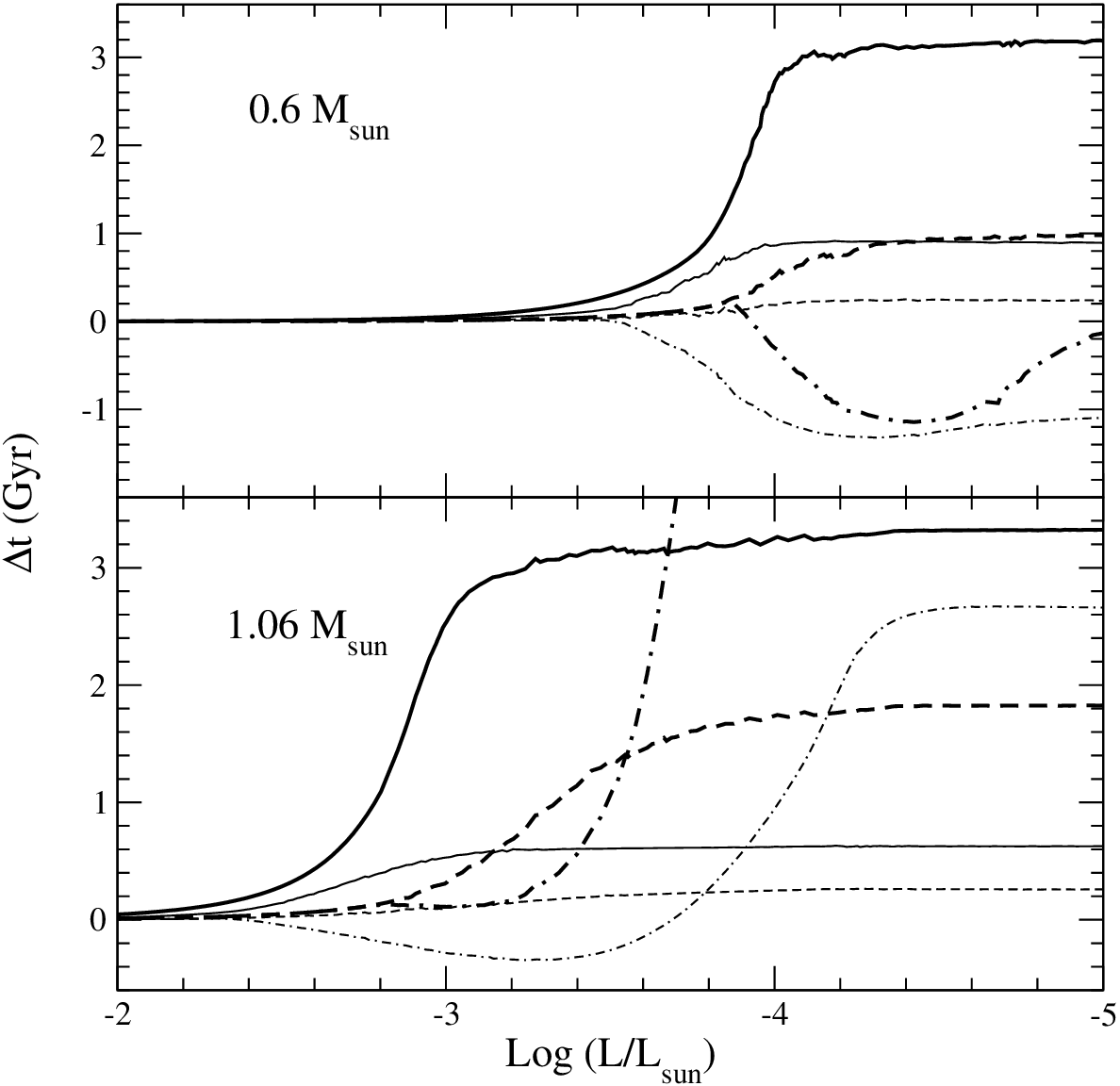}
\caption{Difference  in  evolutionary  times  between  sequences  with
         various assumptions about diffusion and the sequence in which
         diffusion  is neglected. Solid,  dashed and  dot-dashed lines
         correspond to sequences with  $D= 5D_s$, $D= D_s$, and glassy
         state with $D= D_{\rm glass}$.  The thick (thin) set of lines
         are  for  the  model  sequence with  carbon-  (oxygen-)  rich
         cores.}
\label{comparo}
\end{center}
\end{figure}

In closing,  we note that  the more sophisticated  and self-consistent
treatment of white dwarf evolution with diffusively evolving $^{22}$Ne
abundances done  in this work  yields results that are  in qualitative
agreement with those of Deloye  \& Bildsten (2002) which were obtained
using  a global  and much  more  simplified treatment  of white  dwarf
evolution.   However,  there  are  quantitative differences.   In  the
interest  of  comparison,  we  show  in Fig.   \ref{comparo}  the  age
difference between our  sequences with $D= 5D_s$ and  $D= D_s$ and the
sequence in which $^{22}$Ne  diffusion is not considered.  Results for
both  carbon-rich and oxygen-rich  core compositions  are illustrated.
This  figure can  be compared  with figure  10 of  Deloye  \& Bildsten
(2002). Although  a direct  comparison is not  easy because  Deloye \&
Bildsten (2002)  consider cores  composed by a  mixture of  carbon and
oxygen,  a close  inspection  of  the figures  reveals  that for  both
stellar masses  our calculations predict  age differences larger  by a
factor of $\sim 2$ when  compared with the age differences obtained by
Deloye  \& Bildsten  (2002).   In fact,  we  find that  the impact  of
$^{22}$Ne diffusion  on the white dwarf cooling  ages is substantially
more important than originally  inferred by Deloye \& Bildsten (2002).
For  the case  in which  a glassy  transition is  considered,  we find
marked differences between both  sets of calculations for the low-mass
sequence.  Indeed,  in our  calculations, the glassy  sequence remains
younger than  the standard  sequence without diffusion  throughout the
entire evolution (we stress that  the glassy sequence does not release
latent heat), while  in the Deloye \& Bildsten  (2002) treatment, once
the  crystallization process  is  finished, the  released energy  from
$^{22}$Ne diffusion in the glassy sequence yields an age increase.

\section{Discussion and conclusions}

Motivated by the theoretical proposal that the cooling of white dwarfs
could  be altered  by the  sedimentation  of $^{22}$Ne  in the  liquid
interior of these stars (Bravo  1992; Bildsten \& Hall 2001; Deloye \&
Bildsten 2002) and  the possibility that this could  leave imprints in
the white dwarf population of metal-rich systems, we have presented in
this paper  detailed evolutionary calculations to  address this issue.
These are the first fully self-consistent evolutionary calculations of
this effect. We have  explored several possibilities. In particular we
have thoroughly  analyzed two classes of cooling  sequences. The first
class corresponds to an otherwise typical $0.6\,M_{\sun}$ white dwarf,
whereas  the second  one corresponds  to a  massive  $1.06\, M_{\sun}$
white  dwarf. Also  the  effects  of the  core  composition have  been
addressed by following cooling sequences for carbon cores with a small
admixture of  $^{22}$Ne and the  corresponding ones in which  the main
constituent of the core was  oxygen. The sensitivity of our results to
the precise  value of the  rather uncertain diffusion  coefficient has
also been  explored.  Finally,  we have also  analyzed the  effects of
assuming  that  the  white  dwarf  instead  of  crystallizing  at  low
temperatures experiences a transition to  a glassy state.  All in all,
we  have  found  that  the  sedimentation  of  $^{22}$Ne  has  notable
consequences  in the  evolutionary  timescales of  white dwarfs.   The
associated energy release notably  delays the cooling of white dwarfs.
The  precise value  of  the  delay depends  on  the adopted  diffusion
coefficient, on the  mass of the white dwarf,  on the core composition
and on the possibility that  white dwarfs experience a transition to a
glassy state.  The  delay is larger for more  massive white dwarfs and
for  carbon-rich cores.  Specifically,  for the  case of  an efficient
diffusion the  delay introduced in the $1.06  \, M_{\sun}$ carbon-rich
sequence is $\sim 3.2$  Gyr at $\log({L/L_{\sun}}) \approx -4$ whereas
it only  amounts to 0.6 Gyr  in the case of  the oxygen-rich sequence.
In low-mass white dwarfs, appreciable delays in the cooling rates take
place but  only at  much smaller luminosities.   These results  are in
relative good agreement with the results of Deloye \& Bildsten (2002),
although we find that the delays obtained here are somewhat larger.

\begin{figure}
\begin{center}
\includegraphics[clip,width=0.9\columnwidth]{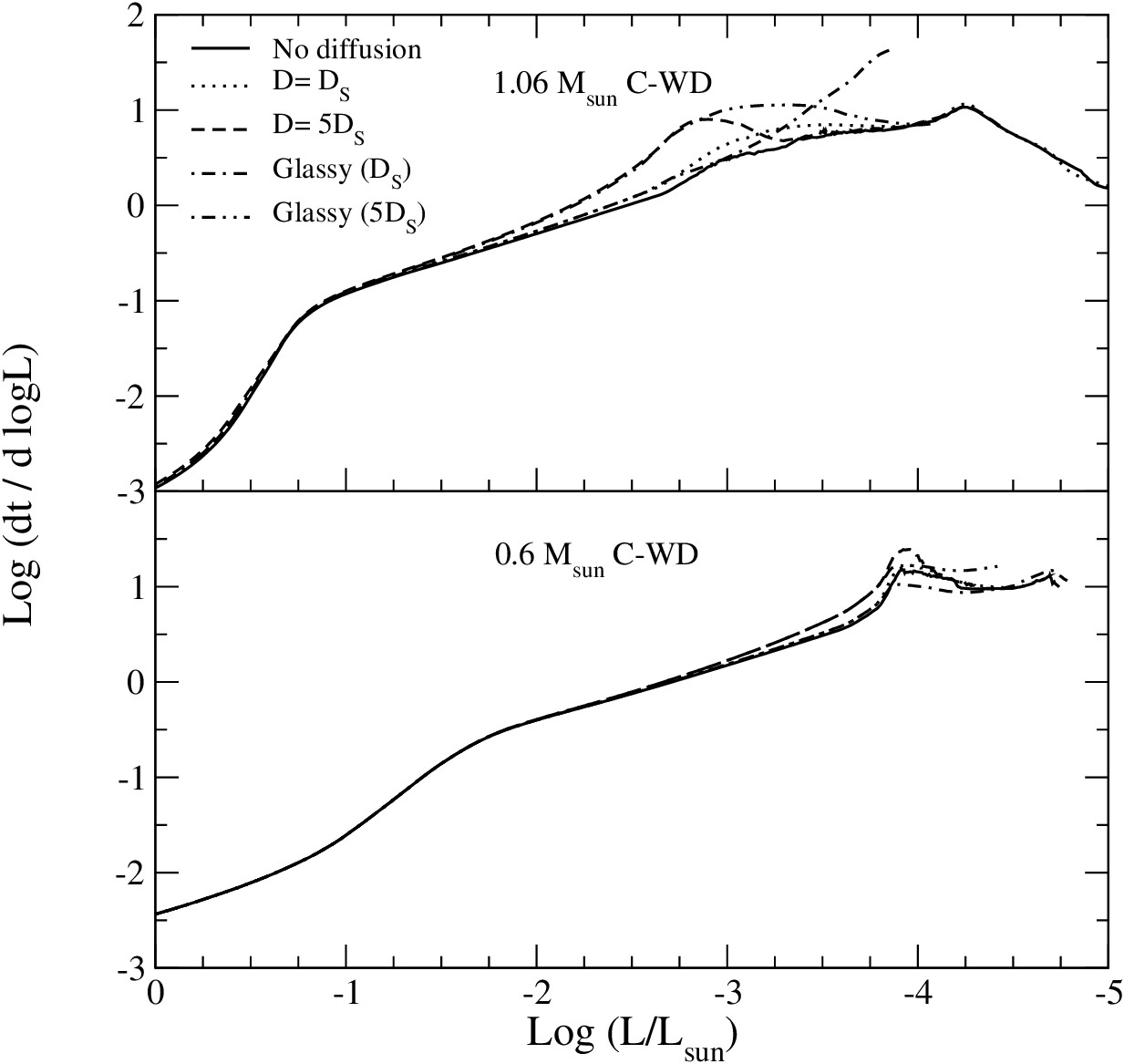}
\caption{The   characteristic   cooling  time   $-{\rm   d}t  /   {\rm
         d}\log(L/L_{\sun})$ for the 1.06  and $0.6 \, M_{\sun}$ white
         dwarf  evolutionary  sequences with  carbon-rich  cores as  a
         function of  luminosity.  The  solid line corresponds  to the
         case  in which  $^{22}$Ne diffusion  is not  considered.  The
         dotted and dashed lines display  the results for $D= D_s$ and
         $D= 5D_s$, while the remaining curves correspond to the cases
         in  which the  models enter  a glassy  state. For  the latter
         situation,  we show results  for $D=  D_{\rm glass}$  and $D=
         5D_{\rm  glass}$.  Only ages  smaller than  14 Gyr  have been
         considered.}
\label{lf_c}
\end{center}
\end{figure}

\begin{figure}
\begin{center}
\includegraphics[clip,width=0.9\columnwidth]{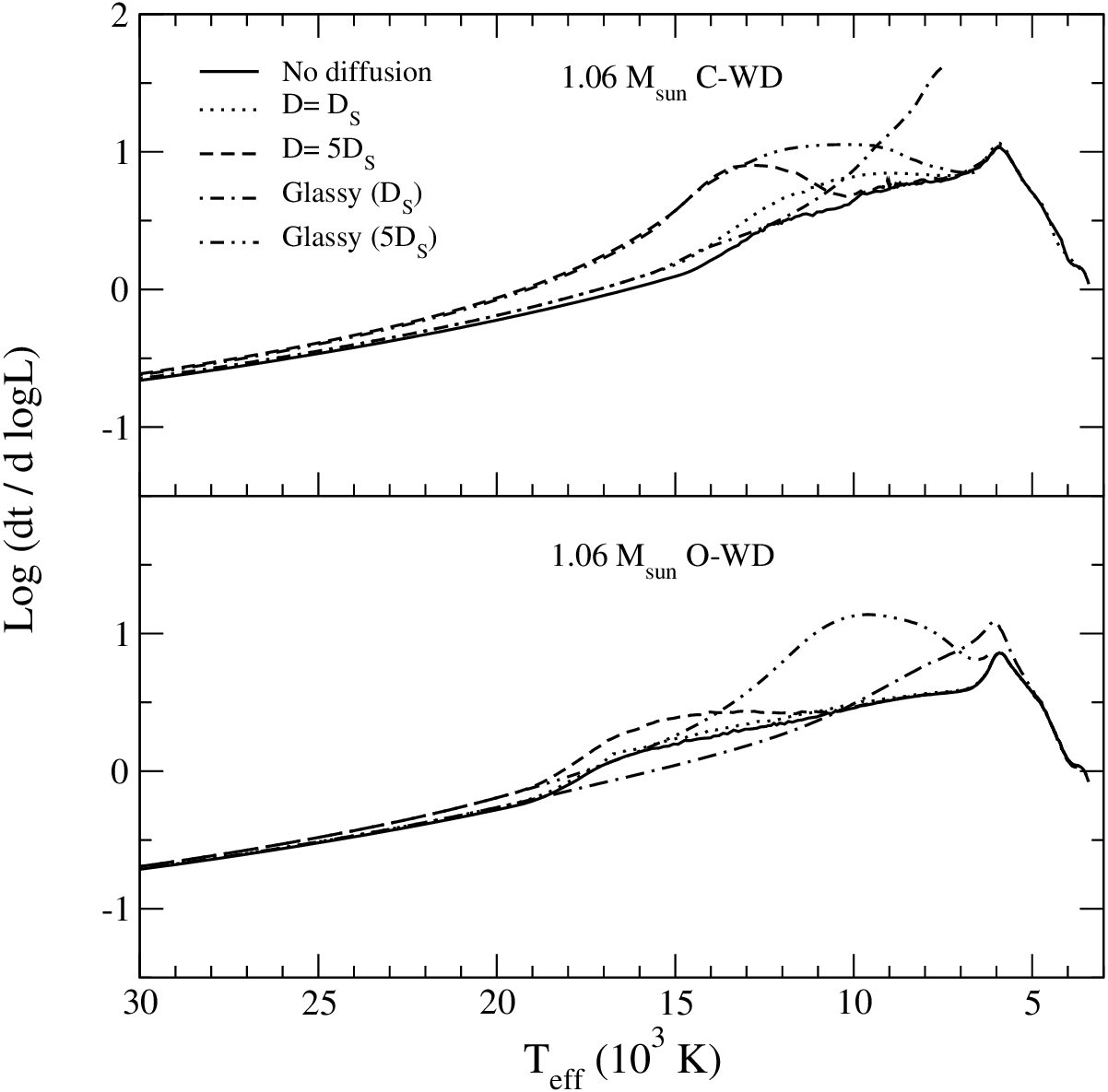}
\caption{Same as Fig.   \ref{lf_c}. Results are now shown  in terms of
         the effective  temperature. The  upper (bottom) panel  is for
         the  $1.06 \,  M_{\sun}$ white  dwarf sequences  with carbon-
         (oxygen-) rich cores.}
\label{lf_teff_106}
\end{center}
\end{figure}

To qualitatively  assess the effect of $^{22}$Ne  sedimentation on the
white dwarf luminosity function, we have derived individual luminosity
functions from  our cooling curves.  The  number of white  dwarfs of a
given  mass is  proportional to  the characteristic  cooling  time, $n
\propto {\rm d}t / {\rm d}\log(L/L_{\sun})$.  We plot this quantity in
Fig.  \ref{lf_c} for the 1.06 and $0.6\, M_{\sun}$ white dwarf cooling
sequences  with  carbon-rich  cores  as  a  function  of  the  surface
luminosity.   The  characteristic  cooling   times  in  terms  of  the
effective temperature  for the $1.06 \, M_{\sun}$  white dwarf cooling
sequences  with  carbon-  and  oxygen-rich  cores are  shown  in  Fig.
\ref{lf_teff_106}. In both figures,  the solid line corresponds to the
case in which $^{22}$Ne diffusion  is not considered and the remaining
curves to  the cases of different diffusion  assumptions, indicated in
the  legend.  At large luminosities the  signature of  the accelerated
evolution  due  to  neutrino  losses  is  clearly  noticeable  in  the
characteristic  cooling  times.   At  the  extreme faint  end  of  the
luminosity range, the sharp decline  in the luminosity function in the
massive sequences reflects the onset of Debye cooling, stage which is
reached at different ages depending on the assumption about diffusion.

Note  that the  effect of  $^{22}$Ne diffusion  in  the characteristic
cooling  times  of  low-mass   white  dwarfs  with  solar  metallicity
progenitors  is   barely  noticeable.    In  sharp  contrast,   it  is
significative in the case of  massive white dwarfs. In fact, depending
on the  internal composition and the uncertainties  weighting upon the
determination of the diffusion  coefficient, an increase in the number
of white  dwarfs by a factor of  $\approx$ 5 could be  expected in the
effective temperature range  where $^{22}$Ne diffusion constitutes the
main  source of stellar  luminosity. In  the case  that $D=  5D_s$ and
assuming carbon-rich  composition, this increase in  the massive white
dwarf population would be expected  at $T_{\rm eff}$ within 12,000 and
16,000 K. For oxygen-rich cores, only a modest increase is expected at
somewhat larger  effective temperatures. If  a transition to  a glassy
state is assumed instead of  crystallization, a marked increase in the
luminosity  function  is  expected  in both  carbon-  and  oxygen-rich
massive  white  dwarfs. In  the  case  of  an oxygen-rich  core,  this
increase  takes  place  at  the effective  temperature  range  between
$\approx$ 12,000 and 8,000  K.  Interestingly enough, a possible large
excess in the  number of massive white dwarfs  has been reported below
$T_{\rm eff}= 12,000$ K (Kleinman  et al.  2004; Liebert et al.  2005;
Kepler et  al. 2007).   Although this increase  is believed to  be the
result  of  a problem  in  the line  fitting  procedure,  it might  be
possible that such  an increase in the number  of massive white dwarfs
could in part be reflecting a  decrease in the cooling rate of massive
white dwarfs,  induced by the  gravitational settling of  $^{22}$Ne in
the core of these white dwarfs.

The impact of $^{22}$Ne sedimentation on the white dwarf structure can
be assessed from  pulsating white dwarfs.  Indeed, as  shown by Deloye
\&  Bildsten (2002),  the abundance  gradient in  $^{22}$Ne  caused by
diffusion  in   the  liquid  regions  leaves  its   signature  in  the
Brunt-V\"ais\"al\"a frequency  which, in turn,  modifies the pulsation
spectrum of ZZ Ceti stars at levels compared with the uncertainties of
the  measurements.  More  interestingly, pulsating  white  dwarfs with
measured rates of  period changes can be used to  assess the impact of
$^{22}$Ne sedimentation on the  white dwarf cooling directly.  Indeed,
the delay in the cooling  times resulting from the $^{22}$Ne diffusion
is expected  to alter  the rate  of change of  the periods  of ZZ~Ceti
stars.  For a quantitative inference  of the possible impact, we write
the rate of change of the  pulsation period of a pulsating white dwarf
as

\begin{equation}
\frac{\dot P}{P}=-a \frac{\dot T}{T}+b \frac{\dot R_\star}{R_\star}
\label{pdot}
\end{equation}

\noindent where $T$ is the  temperature of the isothermal core and $a$
and  $b$ are constants  of order  unity which  depend on  the chemical
composition,  thicknesses  of   the  hydrogen  and  helium  envelopes,
equation of state,  and other ingredients involved in  the modeling of
white dwarfs (Winget et al. 1983).  For DA white dwarfs in the ZZ~Ceti
instability   strip,  the   rate  of   change  due   to  gravitational
contraction,  given by  the  second term  of  the right  hand side  of
Eq. (\ref{pdot}), is usually negligible and, thus, the secular rate of
change of the period is directely  related to the speed of cooling and
it is a positive contribution.   The rate of secular period change has
been  measured for  some pulsating  white dwarfs.   For  instance, for
G117-B15A, an  intermediate mass ZZ~Ceti  star, Kepler et  al.  (2005)
have  derived the  secular variation  of the  main observed  period of
215.2~s, $\dot P = (3.57  \pm 0.82 ) \times 10^{-15}$~s~s$^{-1}$, with
unprecedented accuracy.  Other pulsating  white dwarfs --- like L~19-2
and R~548 ---  have also determinations of the  secular rate of period
change but are not as accurate as that of G117-B15A.  We find that for
the  $1.06  \,  M_{\sun}$  carbon-rich  sequence with  $D=  5D_s$  the
predicted  period  change decreases  by  a  factor  of about  1.7,  as
compared  with the period  change in  the case  that diffusion  is not
considered.  This  result is very  preliminary, and an in  depth study
will be done in a forthcoming publication.

In summary, although  the effect of the sedimentation  of $^{22}$Ne is
sizeable its impact  on the white dwarf luminosity  function should be
minor, except for a modest increase  in the derived ages, for the most
reasonable assumptions. However, the  imprints of the sedimentation of
$^{22}$Ne  could be  detectable using  pulsating white  dwarfs  in the
appropriate  effective temperature  range  with accurately  determined
rates of change of the observed periods.

\acknowledgments

Part of this  work was supported by the  MEC grants AYA05-08013-C03-01
and 02,  by the European  Union FEDER funds,  by the AGAUR and  by PIP
6521 grant from CONICET.

\end{document}